# Foreign Signal Radar

Wei Jiao*

## Abstract

We introduce a new machine learning approach to detect value-relevant foreign information for both domestic and multinational companies. Candidate foreign signals include lagged returns of stock markets and individual stocks across 47 foreign markets. By training over 100,000 models, we capture stock-specific, time-varying relationships between foreign signals and U.S. stock returns. Foreign signals exhibit out-of-sample return predictability for a subset of U.S. stocks across domestic and multinational companies. Valuable foreign signals are not concentrated in those largest foreign markets nor foreign firms in the same industry as U.S. firms. Signal importance analysis reveals the price discovery of foreign information is significantly slower for information from emerging and low-media-coverage markets and among stocks with lower foreign institutional ownership but is accelerated during the COVID-19 crisis. Our study suggests that machine learning-based investment strategies leveraging foreign signals can emerge as important mechanisms to improve the market efficiency of foreign information.

**This version:** March 20, 2025.

---

* I thank seminar participants at Rutgers, the Wolfe research annual global quantitative and macro investment conference, the FMA annual meeting, the EFA annual meeting, the WPI fintech conference, and the Bar Harvard conference for their comments and suggestions. I acknowledge partial financial support from the David Whitcomb Center for Research in Financial Services of Rutgers University. *Correspondence to*: Wei Jiao, Rutgers University. Email: wei.jiao@rutgers.edu. *ORCID*: orcid.org/0000-0002-3998-7602

Firms buy and sell products and services globally. Both multinational and domestic firms compete in a global marketplace and establish cross-border supply chains. Recent years have also seen that wars, pandemics, trade disputes, and other economic and political factors could suddenly reshape a country's business environment. Multiple countries' stock markets dropped over 30% during the COVID-19 outbreak. With the onset of the 2022 Russia-Ukraine war, major American companies pulled businesses out of Russia. Signals that reflect foreign market information are of growing importance to firm valuation. Detecting value-relevant foreign signals and studying foreign information dissemination is critical for understanding capital market efficiency and firm valuations.

Investors and researchers traditionally rely on human intuition and reasoning to identify valuable information. However, prior literature has documented that global business exposure brings challenges for market participants in processing foreign information (e.g., Thomas (1999), Callen, Hope, and Segal (2005) Duru and Reeb (2002), Khurana, Pereira, and Raman (2003), Li, Richardson, and Tuna (2014), Huang (2015), Brochet, Naranjo, and Yu (2016), Finke and Weigert (2017), Lundholm, Rahman, and Rogo (2018), and Fang and Lin (2025)). And these existing studies about foreign information typically focus on multinational firms since it is relatively straightforward to identify signals related to multinationals' revenues in foreign markets.

Rather than using economic intuition, we take a different approach and rely on machine learning algorithms as the "Radar" to scan the pool of candidate foreign signals and detect valuable ones. We then use human intuition to examine whether the signals identified by algorithms are economically meaningful. Machine learning algorithms are designed to maximize out-of-sample predictive accuracy by modeling the relationships between predictors and the target of interest. Their ability to curb overfitting helps these algorithms identify valuable predictors and discard extraneous ones.[1] Our method provides an approach to detect value-relevant foreign signals for both multinational and domestic companies.

Specifically, we search for foreign signals that can forecast the daily returns of the S&P 500 constituents. Our primary pool of 188 (47 × 4) candidate foreign signals includes the lagged weekly stock market returns in 47 non-U.S. markets during the previous four weeks. We also challenge machine learning algorithms with a substantially larger pool of ≈13,000 candidate foreign signals. This extensive pool encompasses the lagged weekly individual stock returns and stock market returns in 47 non-U.S. markets. The stock market and individual stock returns in 47 non-U.S. markets are denominated in local currency. We employ multiple machine learning algorithms: least absolute shrinkage and selection operator (LASSO) regression, random forest (RF), gradient boosted tree (GB), and neural network (NN).

Our approach differs from most prior studies applying machine learning for predictions, which typically train one model at a time to generate predictions for all firms. The intricate channels through which foreign signals impact U.S. stock returns necessitate stock-specific time-varying models. For instance, the 2022 Russia-Ukraine war caused the Russian stock market to drop 30% within a single month. This conflict negatively impacted many U.S. companies selling consumer goods in Russia but was positive news for defense contractors. Figure 1, for example, shows after the outbreak of the war, Lockheed Martin's stock price rose significantly, whereas the stock prices of McDonald's and Starbucks declined.

To capture such relationships, we trained over 100,000 models. Each model is trained on one U.S. stock's daily observations in rolling four-quarter estimation windows (≈252 daily observations). We then use each trained model to make out-of-sample forecasts for this stock's daily returns. Finally, we repeat the procedures to obtain out-of-sample return forecasts for all S&P 500 stocks. In this study, we focus on the S&P 500 stocks because of the high computational demands to train models for each stock-quarter using multiple machine learning algorithms.

---

[1] In contrast, traditional ordinary least squares (OLS) regressions maximize in-sample estimation accuracy. For comparison, we also apply OLS regressions to identify foreign signals and present the findings in this study.

We first examine the out-of-sample predictive power by the pool of 188 foreign stock market return signals. We start by constructing long-short portfolios using forecasted daily returns by foreign signals. Given the limited cross-section of about 500 stocks, we first look at the stocks with top or bottom 5% predicted returns. Each day, the portfolio buys stocks with top 5% predicted returns and sells short those with bottom 5% predicted returns. The long-short portfolio constructed based on LASSO's predicted returns generates abnormal returns of 5.63 basis points per day or 14.2% annually. Based on the predicted returns by RF, GB, and NN, the long-short portfolio generates abnormal returns of 3.81, 3.80, and 5.77 basis points per day, respectively. We observe the abnormal returns for the early and later periods, adjusting the U.S. or global risk factors, or aggregating daily returns to monthly returns.

We also find that the performance of the long-short portfolio concentrates on periods with low media coverage of foreign news. A large amount of media coverage of foreign news can draw investor attention to foreign information, accelerate the dissemination of foreign information into stock prices, and leave less room for the return predictability of foreign signals.

We then sort stocks into deciles based on daily return forecasts and report the abnormal returns for portfolios in each decile and for the long-short decile spread portfolios. The long-short decile spread portfolios yield positive and significant abnormal returns. Unlike traditional firm characteristics' impact on returns, the performance of these portfolios formed on return forecasts by foreign signals does not decrease monotonically from the top decile portfolio to the bottom one.

This finding suggests that a subset of U.S. stocks are detected to be materially exposed to foreign information.

The out-of-sample $R^2_{oos}$ computed using predicted and realized daily returns also reveal that our foreign signals have return predictability for a fraction of U.S. stocks. The predicted returns by LASSO, RF, GB, and NN exhibit positive out-of-sample predictive $R^2_{oos}$ for 24%, 22%, 36%, and 8% of the S&P 500 stocks, respectively. Among the stocks with positive $R^2_{oos}$, the average $R^2_{oos}$ is 1.6%, 3.0%, 2.0%, and 3.6% based on the forecasts by LASSO, RF, GB, and NN. Importantly, these statistics regarding $R^2_{oos}$ show remarkable similarity across multinational and domestic firms and firms in different industries. This finding indicates that the impact of foreign signals on stock returns extends across *multinational and domestic firms and industries*.

To unpack the "black boxes" of machine learning models, we analyze signals' importance and contribution to models' return forecasts. The signal importance analyses are based on stock-quarters with positive out-of-sample $R^2_{oos}$, which suggests the models capture some meaningful relationships between foreign signals and subsequent stock returns. Our analyses overall suggest that foreign signals identified by machine learning models are economically meaningful.

Prior studies show that there are delays in incorporating foreign information into multinationals' prices due to investors' inattention and limited understanding of foreign information.[2] Our approach allows us to study both domestic and multinational firms and shows that signal importance decreases gradually from lagged 1-week to lagged 4-week foreign signals. Namely, U.S. stock prices have absorbed the information in the lagged 1-week signals to a much lesser extent than in lagged 4-week signals. This finding indicates the information embedded in foreign signals is slowly disseminated into both domestic and multinational firms' stock prices.

---

[2] For example, see Huang (2015) and Finke and Weigert (2017).

The estimations based on LASSO models indicate that foreign information takes four weeks to be fully incorporated into U.S. stock prices. The estimated time window is five weeks for RF models, six weeks for GB models, and up to eight weeks for NN models.

The heightened visibility of foreign news in the media can help accelerate the dissemination of foreign information. By analyzing how fast the importance decreases from lagged 1-week to lagged 4-week signals, we find that U.S. stock prices take much longer to incorporate the information in signals from foreign markets with low media coverage. Emerging markets could exhibit poor quality and low credibility of financial information and present a high level of information asymmetries (e.g., Ahearne, Griever, and Warnock (2004) and Kho, Stulz, and Warnock (2009)). Accordingly, we find that the price discovery is much slower for information from emerging markets.

Foreign institutional investors can possess advantages in gathering information from their home countries and accelerate the incorporation of foreign information into stock prices (e.g., He et al. (2013) and Kacperczyk, Sundaresan, and Wang (2021)). We find that U.S. stocks with lower foreign institutional ownership take a longer time to absorb the information in foreign signals. The overall institutional ownership, however, does not impact the price discovery of information in foreign signals after we control foreign institutional ownership.

The COVID-19 pandemic drove millions of people out of work and caused economic shutdowns in many countries. Extensive news coverage of outbreaks in different countries, coupled with substantial fluctuations in foreign stock markets, could draw heightened investor attention. Accordingly, we find that the dissemination of foreign information was notably accelerated during the COVID-19 outbreak.

The price informativeness of the S&P 500 stocks has increased steadily since 1960 (Bai, Philippon, and Savov (2016)). While we document the slow diffusion of foreign information, our results also reveal that over time, financial markets have become more efficient in incorporating foreign information into asset prices, and the gain is mainly from the improved price discovery of emerging-market information.

An ordinary investor might conjecture that the most important markets are those with the largest gross domestic product or stock market capitalization. Our analysis shows valuable foreign signals are not concentrated in the largest countries. Yet, for example, Qatar frequently emerges as one of the five most important markets. Qatar's importance is possibly due to its critical role in the global natural gas and oil supply, and Qatar often acts as a key mediator in the geopolitical conflicts in the Middle East.

Finally, we challenge machine learning algorithms with a substantially larger pool of ≈13,000 candidate foreign signals, including weekly individual stock and stock market returns in 47 non-U.S. markets over the preceding four weeks. All the stocks with a top 10% market capitalization in each market are included. Machine learning algorithms can still detect valuable foreign signals from this large pool. When we use human intuition to identify foreign signals, it seems natural to prioritize foreign firms within the same industry as U.S. firms. However, our analysis suggests that valuable foreign signals are not concentrated in foreign firms in the same industries as U.S. firms. In addition, we find that this extensive pool does not improve the overall forecast accuracy due to the increased sparsity of useful foreign signals in the pool.

Our study first contributes to the literature on the challenges and frictions that investors and financial analysts face when processing foreign information.[3] Rather than solely using human intuition to identify value-relevant foreign information, we contribute a new approach that harnesses machine learning algorithms to detect valuable foreign signals. Afterward, market participants can apply their reasoning to interpret the economic implications of the identified signals. Moreover, existing studies predominantly examine the difficulties of processing foreign information for multinational firms. In this paper, we demonstrate that foreign information significantly influences the valuation of both domestic and multinational companies and takes multiple weeks to be incorporated into their asset prices.

Our work also contributes to the growing literature on the transformative role of machine learning and artificial intelligence in the disclosure, processing, and dissemination of information in capital markets (e.g., Costello, Down, and Mehta (2020), Hsieh et al. (2020), Erel, Stern, Tan, and Weisbach (2021), Goldstein, Spatt, and Ye (2021), Li, Mai, Shen, and Yan (2021), Hallman, Kartapanis, and Schmidt (2022), Cao, Jiang, Yang, and Zhang (2023), Guenther et al. (2023), and Drake, Moon, and Warren (2024)). Our findings highlight that machine learning-based investment strategies exploiting foreign signals can emerge as indispensable mechanisms for the price discovery of foreign information.

Lastly, our study broadens the application of machine learning in analyzing asset prices and investment decisions. Existing studies primarily use machine learning algorithms to choose between firm characteristics and signals already identified by other researchers or model new relationships between these already-known signals and asset returns.[4] In contrast, this paper is tackling a problem where prior knowledge about which foreign signal in the pool would work mostly does not exist. We contribute to the literature by documenting that foreign signals identified through machine learning can exhibit out-of-sample return predictability, thereby enhancing the potential for more informed and dynamic investment decision-making.

---

[3] For example, see Thomas (1999), Callen, Hope, and Segal (2005) Duru and Reeb (2002), Khurana, Pereira, and Raman (2003), Li, Richardson, and Tuna (2014), Huang (2015), Brochet, Naranjo, and Yu (2016), Finke and Weigert (2017), Lundholm, Rahman, and Rogo (2018), and Fang and Lin (2025).

**1. Model and Data**

*1.1 Model training and predicting procedures*

The intricate channels through which foreign signals impact U.S. stock returns necessitate stock-specific time-varying models. For instance, the 2022 Russia-Ukraine war caused the Russian stock market to drop 30% within a single month. This conflict negatively impacted many U.S. companies selling consumer goods in Russia but was positive news for defense contractors. Figure 1, for example, shows after the outbreak of the War, Lockheed Martin's stock price rose significantly, whereas the stock prices of McDonald's and Starbucks declined. The extent of the War's impact also varied depending on companies' exposure to the Russian markets. After the outbreak of the War, major American companies pulled businesses out of Russia. Consequently, the relationships between Russian signals and the returns of those companies that exited Russia would change significantly.

Figure 2 illustrates the procedures to train models and generate return forecasts. We train a model *m* using one machine learning algorithm based on daily observations from quarter *T-3* to quarter *T*. For each observation in the training data, the U.S. stock *i*'s return on that day is linked with lagged foreign signals. Our candidate foreign signals are the leading stock market index returns in 47 non-U.S. markets. We use lagged weekly returns of these stock market indices during the previous four weeks, which results in a pool of 188 (47 × 4) foreign signals. The content in the box surrounded by dotted lines exemplifies the lagged foreign signals for one daily observation. The trained model *m* thus captures the relationships between lagged foreign signals and U.S. stock *i*'s daily returns during the training period.

---

[4] For example, see Freyberger, Neuhierl, and Weber (2020), Gu, Kelly, and Xiu (2020), Kozak, Nagel, and Santosh (2020), Leippold, Wang, and Zhou (2022), Bryzgalova, Pelger, and Zhu (2023), Cong et al. (2023), and Hellum, Pedersen, and Rønn-Nielsen (2023).

To forecast U.S. stock *i*'s returns on each trading day in quarter *T+1*, we feed the trained model *m* with the corresponding lagged weekly foreign market returns. The outputs of model *m* are U.S. stock *i*'s out-of-sample predicted returns on each trading day in quarter *T+1*. When repeating the steps for all the quarters in the sample, we obtain U.S. stock *i*'s return forecasts for all the trading days in the sample. We then implement the whole procedure for each stock in our sample to generate return forecasts for all the stocks. This way, we build stock-quarter-specific models capturing quarterly updated relationships between lagged foreign signals and each U.S. stock's daily returns.

For each stock, we choose to train models at quarterly frequency, because it helps capture the time-varying relationships and lower the computational demands. Investors with adequate computing power can train models at a daily frequency to implement trading strategies based on foreign signals.

*1.2 Data description*

Due to the high computational demands to train models for each stock-quarter using multiple machine learning algorithms, we focus on forecasting returns of the S&P 500 stocks on every trading day from 2012Q1 to 2022Q1. We obtain U.S. stock returns and the S&P 500 index returns from the Center for Research on Security Prices (CRSP). The data on leading stock market index returns and individual stock returns in 47 non-U.S. markets are from Thomson Reuters

Datastream International.[5] The list of the 47 leading stock market indices is in the Internet appendix II. The returns of stock markets and individual stocks in 47 non-U.S. markets are in local currency to remove the impact of exchange rate fluctuations.

The geographical sale information is from Compustat Segment files. Data on risk factors, U.S. Treasury-bill yield, and industry classification are obtained from Kenneth French's data library. The industry information for foreign stocks is from Worldscope. The data on institutional ownership are obtained from the Thomson Reuters Institutional Holdings (13F) Database. The data on U.S. trade in goods with foreign markets are from the U.S. Census Bureau.

*1.3 Machine learning algorithms*

Traditional ordinary least squares (OLS) regressions estimate models by minimizing in-sample predictive errors, which could lead to poor out-of-sample predictive performance because of overfitting. In contrast, machine learning algorithms are designed to minimize out-of-sample predictive errors. We employ four machine learning algorithms that are good at handling tabular data and are commonly used in the literature: least absolute shrinkage and selection operator (LASSO) regression, random forest (RF), gradient boosted tree (GB), and neural network (NN). RF, GB, and NN are also not constrained by the linear structure and parametric assumptions. In our study, we train models based on these algorithms to minimize the mean squared prediction error as the objective function.

LASSO mitigates overfitting and improves predictive accuracy through L1 regularization, which suppresses weak predictors' coefficients to zero and generates parsimonious models. RF is an ensemble algorithm that fits many decision trees on various sub-samples of the data and sub-sets of signals and uses averaging to improve predictive accuracy and control overfitting. GB is also a tree-based ensemble algorithm. It curbs overfitting by training a sequence of decision trees on various sub-samples of the data and sub-sets of signals. Each tree in GB gradually improves predictive accuracy by learning the previous tree's errors.

---

[5] These 47 markets are the ones included in the MSCI all country world index (MSCI ACWI).

For NN, we focus on the traditional feedforward network, a multi-layer collection of mathematical neurons that connect and communicate with each other. The network includes an input layer of multiple neurons receiving initial candidate predictors. Multiple hidden layers with neurons in the network model the relationships between their own inputs and outputs. Non-linear activation functions in each neuron introduce nonlinearity. The output layer in the network aggregates hidden layers' outputs into a final prediction. We also apply L1 kernel regularization in NN, which shrinks the weights on weak inputs to zero, to mitigate overfitting.

LASSO, RF, and GB are implemented using *Scikit-learn*, a widely used machine learning library.[6] We train NN using *Tensorflow*, a software library for machine learning and artificial intelligence.[7] Details on hyperparameter tuning are in the Internet Appendix III. We tune hyperparameters using *Optuna* (Akiba et al. (2019)), an advanced hyperparameter optimization framework to automate hyperparameter search.[8]

LASSO applies the L1 regularization to select signals, which preserves signals with strong coefficients and shrinks the coefficients on signals with weak covariance with the target variable to zero. The scale of signals can influence the magnitude of coefficient estimates and hence impact the estimated models based on LASSO. In our sample, the scale of weekly leading stock market return signals from different countries varies greatly. For example, Greece's lagged 1-week market returns have a standard deviation of 4.5%, while the standard deviation of Malaysia's lagged 1-

---

[6] For details about Scikit-learn, please see https://scikit-learn.org/stable/index.html
[7] For details about Tensorflow, please see https://www.tensorflow.org/
[8] For details about Optuna, please see https://optuna.org/

week market returns is only 1.5%. Therefore, we standardize our foreign signals to make each of them have zero mean and a standard deviation of one. We also use standardization for NN models because L1 kernel regularization is included in NN to curb the overfitting problem.[9]

We use the above four algorithms plus OLS regressions and each of the two foreign signal pools to train models following the procedures in subsection 1.1. To the end, we trained over 100,000 models to forecast the S&P 500 stock returns on each trading day in the sample period.

## 2. Performance of Portfolios Formed on Return Forecasts by Foreign Signals

In this section, we examine the performance of portfolios formed on return forecasts by the pool of 188 foreign stock market return signals. If machine learning algorithms can identify foreign signals to predict stock returns, we should expect portfolios formed on these return forecasts to deliver superior out-of-sample performance.

*2.1 Performance of stocks with top or bottom return forecasts*

In Panel A of Table 1, we rank stocks based on their return forecasts and report daily portfolio performance. Given the limited cross-section of about 500 stocks in our sample, we first look at the stocks with top or bottom 5% predicted returns. In Column (1), we construct a portfolio consisting of stocks with top 5% forecasted returns by LASSO and equally weight each stock. The average return of this portfolio in excess of the daily one-month U.S. Treasury yield is 10.40 basis points per day with a t-statistic of 3.44.

We also analyze the risk-adjusted daily returns of the portfolio. We regress daily excess returns of the portfolio on the market (MKT), size (SMB), value (HML), momentum (MOM), profitability (RMW), and investment (CMA) factors (Fama and French (1993, 2015) and Carhart (1997)) to compute the six-factor alpha. After controlling these risk factors, the six-factor alpha is 3.69 basis points per day with a t-statistic of 2.64. This portfolio shows significantly positive loadings on market, size, and value factors and negative loadings on the momentum factor. [10]

---
[9] For details about regularization, please see
https://www.tensorflow.org/api_docs/python/tf/keras/regularizers/Regularizer

In Column (2), the portfolio consists of stocks with bottom 5% forecasted returns by LASSO. We equally weight each stock in the portfolio. The average return of this portfolio in excess of the daily one-month U.S. Treasury yield is 4.83 basis points per day with a t-statistic of 1.41. The six-factor alpha is -1.94 basis points per day with a t-statistic of -1.36. This portfolio shows significantly positive loadings on market, size, value, and investment factors and negative loadings on momentum and profitability factors.

In Column (3), we construct a long-short portfolio that buys stocks with top 5% forecasted returns and sells short stocks with bottom 5% forecasted returns. The average return of this long-short portfolio is 5.57 basis points per day with a t-statistic of 2.43. The six-factor alpha of this long-short portfolio is 5.63 basis points per day with a t-statistic of 2.77. This portfolio shows significantly positive loadings on value and momentum factors.

In Columns (4) to (6), we use similar procedures to construct portfolios based on return forecasts by RF. The six-alpha of the portfolio consisting of stocks with top 5% forecasted returns is 3.22 basis points per day with a t-statistic of 3.05. The six-factor alpha of the portfolio consisting of stocks with bottom 5% forecasted returns is -0.59 basis points per day with a t-statistic of -0.51. The long-short portfolio delivers a six-factor alpha of 3.81 with a t-statistic of 2.51.

---

[10] The objective function in the training process is the mean squared prediction errors, which equally weight the forecast errors of each stock. Thus, it's natural to study equal-weighted portfolios in our analysis. Our sample focuses on the S&P 500 stocks, so illiquidity and trading costs should not be significant concerns in forming portfolios. We discuss the impact of trading costs in section 2.4. We also report the value-weighted portfolio performance in the robustness check section 7.

In Columns (7) to (9), we construct portfolios based on return forecasts by GB. The six-alpha of the portfolio of stocks with top 5% forecasted returns is 3.23 basis points per day with a t-statistic of 3.14. The six-factor alpha of the portfolio of stocks with bottom 5% forecasted returns is -0.57 basis points per day with a t-statistic of -0.44. The long-short portfolio generates the six-factor alpha of 3.80 with a t-statistic of 2.34.

In Columns (10) to (12), we present the findings for portfolios constructed using return forecasts by NN. Stocks with top 5% forecasted returns generate the six-factor alpha of 3.28 basis points per day with a t-statistic of 2.77. The six-factor alpha of the portfolio of stocks with bottom 5% forecasted returns is -2.49 basis points per day with a t-statistic of -1.79. And the long-short portfolio shows a six-factor alpha of 5.77 basis points per day with a t-statistic of 3.09.

We also create combined portfolios by equally weighting the daily returns of the four portfolios consisting of stocks with top (bottom) 5% forecasted returns by LASSO, RF, GB, and NN. Column (13) shows the six-factor alpha of the combined portfolio for stocks with top 5% forecasted returns is 3.36 basis points per day with a t-statistic of 3.87. Column (14) shows the six-factor alpha of the combined portfolio for stocks with bottom 5% forecasted returns is -1.40 basis points per day with a t-statistic of -1.34. Column (15)'s long-short portfolio buys the portfolio of Column (13) and sells short the portfolio of Column (14). This long-short portfolio generates a six-factor alpha of 4.75 basis points per day with a t-statistic of 3.80.

As a comparison, we form portfolios using the return forecasts by OLS regressions. In Columns (16) to (18), we find that none of the three portfolios generate statistically significant abnormal returns. In fact, the portfolio of stocks with top 5% forecasted returns by OLS delivers even lower average daily returns than the portfolio of stocks with bottom 5% forecasted returns.

On balance, these findings provide evidence that all four machine learning algorithms can detect foreign signals that forecast U.S. stock returns.

*2.2 Foreign news coverage and portfolio performance*

In this subsection, we test whether media coverage of foreign news impacts the performance of portfolios formed on return forecasts by foreign signals. A large amount of media coverage of non-U.S. market news can draw investor attention to foreign information, accelerate the incorporation of foreign information into stock prices, and leave less room for the return predictability of foreign signals.

We obtain the news coverage data from RavenPack, which has been widely used by prior literature.[11] We use the RavenPack Dow Jones and PR edition, which collects worldwide news stories from Dow Jones Newswires, the Wall Street Journal, Barron's, MarketWatch, PRNewswire, Canadian News Wire, LSE Regulatory News Service, and other leading global media organizations. The sources of news stories are primarily in English. For the entities involved in each news story, RavenPack labels their corresponding countries. RavenPack also provides a relevance score for each entity involved in a news story. We only consider those entities with an above 90 relevance score, indicating those entities are strongly relevant in the news stories and usually referenced in the main title or headline of the news items.

In Panel B of Table 1, we separate all the trading days into those with high or low media coverage of foreign news. Media coverage of foreign news is the number of news stories regarding entities in all 47 foreign markets during the four weeks before a trading day, scaled by the total number of news stories covered by RavenPack in the same period. We choose the four weeks before a trading day because we use foreign signals during the previous weeks before a trading day to make the return predictions. Scaling by the total number of news stories in the same period helps mitigate the trend in overall media coverage over time. High (low) media coverage of foreign news indicates those trading days with above (below)-median foreign news coverage during the previous four weeks.

---

[11] For example, see Boudoukh et al. (2019), Chinco, Clark-Joseph, and Ye (2019), Ben-Rephael et al. (2021), Da, Huang, and Jin (2021), and Bushman and Pinto (2024).

The return predictability of foreign signals is much stronger when there is low media coverage of foreign news. Columns (5) and (6) of Panel B, Table 1 shows the superior performance of the long-short portfolio that buys stocks with top 5% forecasted returns and sells short stocks with bottom 5% forecasted returns concentrates on periods with low media coverage of foreign news. Columns (1) and (2) indicate the portfolios of stocks with top 5% forecasted returns also yield much higher returns during periods with low foreign news coverage.

In Panel A of Table 1, we do not find the portfolios of stocks with bottom 5% forecasted returns generate significantly negative six-factor alphas. Columns (3) and (4) of Panel B, however, show that such portfolios tend to yield negative abnormal returns during periods with low media coverage of foreign news. Overall, the findings in this subsection suggest that the return predictability of our foreign signals is more pronounced during periods with low media coverage of foreign news. These findings also provide evidence that our machine learning models capture the impact of foreign information on U.S. stock prices.

*2.3 Trading costs*

Our portfolio formation requires trading at a daily frequency, which can result in substantial trading costs. In this subsection, we assess the performance of stocks with top 5% return forecasts after trading costs.

Frazzini, Israel, and Moskowitz (2015) document that for a large institutional investor, the median trading costs are 6.24 (6.16) basis points at each rebalance for stocks on NYSE (Nasdaq) between August 1998 and September 2013. 6.24 basis points mean that the round-trip trading costs are $6.24\times10^{-4}$ when we rebalance $1 worth of stocks. Our strategy involves only the S&P 500 stocks, which should have much lower trading costs than 6.24 basis points. To have a conservative estimation, we use 6.24 basis points to estimate the after-cost performance.

The average daily turnover for the portfolio of stocks with top 5% return forecasts by LASSO is 32.8%. That means we need to rebalance, on average, $0.328 per day for every $1 investment. Thus, the estimated average daily trading costs would be 2.05 basis points (6.24 basis points × 32.8%). Column (1) of Table 1 Panel A shows this portfolio has a 10.40 basis point per day average return in excess of the daily one-month U.S. Treasury yield. Thus, this portfolio would deliver an average return of 8.35 basis points per day (10.40 – 2.05), net of trading costs.

For the portfolio of stocks with top 5% return forecasts by RF, the average daily trading costs are 4.01 basis points (6.24 basis points × 64.2%). Column (4) of Table 1 Panel A shows this portfolio has a 9.66 basis point per day average excess return. This implies an average daily return of 5.65 basis points net of trading costs. For the portfolio of stocks with top 5% return forecasts by GB, the average daily trading costs are 3.48 basis points (6.24 basis points × 55.8%). Column (7) of Table 1 Panel A shows this portfolio has a 9.72 basis point per day average excess return. Thus, the average daily return after trading costs is estimated at 6.24 basis points. The portfolio of stocks with top 5% return forecasts by NN has average daily trading costs of 2.52 basis points (6.24 basis points × 40.4%), indicating an average daily return after trading costs of 7.18 basis points. The combined portfolio has an average daily turnover of 42.2%, which results in average daily trading costs of 2.65 basis points. Column (13) of Table 1 Panel A shows the combined portfolio has a 9.87 basis point per day average excess return. Therefore, its average return after trading cost would be 7.22 basis points per day or 18.19% annually.

In sum, these analyses provide some evidence that large institutional investors could possibly earn returns using trading strategies based on return forecasts by foreign signals.

*2.4 Sharpe ratio, maximum one-quarter loss, and turnover*

Panel A of Table 1 shows that the abnormal returns concentrate on those stocks with top 5% return forecasts. In this subsection, we describe the Sharpe ratio, maximum one-quarter loss, and portfolio turnover for the portfolio of these stocks. Table A1 of the Internet Appendix I shows the four portfolios of stocks with top 5% forecasted returns by LASSO, RF, GB, and NN have annualized Sharpe ratios over 1. The combined portfolio has the highest Sharpe ratio of 1.19. In comparison, the Sharpe ratio of the S&P 500 index is 0.81, and the Sharpe ratio of the portfolio of stocks with top 5% forecasted returns by OLS is only 0.65.

Maximum one-quarter loss is the most extreme negative return in a quarter. The portfolios of stocks with top 5% forecasts by machine learning algorithms at most deliver around -25% return in a quarter, which is worse than the S&P 500's -20.3%. But the portfolio of stocks with top 5% forecasted returns by OLS has the maximum one-quarter loss at -43.4%.

Turnover describes the percentage of portfolio assets that are changed in a day.[12] Table A1 shows the portfolio based on LASSO's predictions has the lowest average turnover at 32.8% per day. The portfolio based on RF's predictions has the highest average turnover at 64.2% per day. The combined portfolio has an average turnover of 42.2% per day.

*2.5 Performance of stocks in each decile*

In Table 2, we sort stocks into deciles based on their daily return forecasts. We form 10 portfolios by equally weighting stocks in each decile. We report the six-factor alpha for portfolios in each decile and for the long-short decile spread portfolios that buy stocks in the top decile and sell short stocks in the bottom decile.

---

[12] Following Gu, Kelly, and Xiu (2020), we compute turnover in a day for the long-only portfolio as $\frac{1}{2}\sum_i \left| w_{i,t} - \frac{w_{i,t-1}(1+r_{i,t})}{\sum_i w_{i,t-1}(1+r_{i,t})} \right|$, where $w_{i,t-1}$ is the portfolio weight of stock *i* in day *t* and $r_{i,t}$ is stock *i*'s return in day *t*.

For RF, GB, and NN, the top decile and long-short decile spread portfolios yield positive and significant abnormal returns. For example, the six-factor alpha of the long-short decile spread portfolio based on forecasts by NN is 3.95 basis points per day. For LASSO, neither the portfolio of stocks in the top decile nor the long-short decile spread portfolio generates statistically significant abnormal returns. The combined portfolios, which equally weight the returns of portfolios based on LASSO, RF, GB, and NN in each decile, also exhibit positive and significant abnormal returns in the top decile and the long-short decile spread.

In addition, unlike the impact of many traditional firm characteristics on returns, the six-factor alphas here do not decrease monotonically from the top decile portfolio to the bottom one. This finding suggests that a subset of U.S. stocks are detected to be materially exposed to foreign signals. Finally, when we sort stocks into deciles based on the return forecasts by OLS, none of the ten portfolios generate significant abnormal returns.

## 3. Out-of-sample $R^2_{oos}$

The previous section shows that portfolios formed on return forecasts by foreign signals generate abnormal returns. In this section, we measure to what extent the out-of-sample predicted returns by foreign signals can explain the variations in daily U.S. stock returns. The results are based on the return forecasts by the pool of 188 foreign stock market return signals. We calculate the out-of-sample $R^2_{oos}$ following Gu, Kelly, and Xiu (2020), as

$$R^2_{oos} = 1 - \frac{\sum(r_{i,t} - r_{predict,i,t})^2}{\sum(r_{i,t})^2},$$

where $r_{i,t}$ is U.S. stock $i$'s return on day $t$ and $r_{predict,i,t}$ is the predicted return of stock $i$ on day $t$.

We train stock-specific and quarterly updated models to predict one U.S. stock's daily returns in each quarter. Models' performance can vary across stocks and over time. More importantly, even if our models perfectly capture the underlying relationships between foreign signals and returns, the real exposures to foreign signals are also stock-specific and time-varying. Computing $R_{oos}^2$ by pooling all firm-day observations would mask the impact of foreign signals on U.S. stock returns and the performance of our models. Therefore, we compute $R_{oos}^2$ using stock $i$'s real and predicted returns on every trading day in one quarter. For each stock-quarter, we get one $R_{oos}^2$, which reflects the fraction of U.S. stock $i$'s daily return variations in a specific quarter that the return forecasts by foreign signals can explain.

*3.1 Summary statistics of $R_{oos}^2$*

In Table 3, we first report the fraction of the S&P 500 stocks with positive out-of-sample $R_{oos}^2$ in a quarter. When we compute $R_{oos}^2$ using LASSO predictions, on average, 23.5% of the stocks have positive $R_{oos}^2$ in a quarter. When computing $R_{oos}^2$ using RF predictions, on average, 21.6% of the stocks have positive $R_{oos}^2$ in a quarter. Forecasts by GB exhibit positive predictability for 35.8% of the stocks, while forecasts by NN yield positive predictability for 7.7% of the stocks.

Together, we can obtain positive $R_{oos}^2$ for 56.2% of the S&P 500 stocks based on return forecasts by at least one of the four algorithms. Namely, foreign signals show return predictability for about half of the S&P 500 stocks. In contrast, $R_{oos}^2$ computed using predictions by OLS are negative for all stocks. We also present the distribution of $R_{oos}^2$ in Table 3. We observe substantial $R_{oos}^2$ for certain stocks. For example, at the 90$^{th}$ percentile, $R_{oos}^2$ based on return forecasts by RF is 2.6%. At the 99$^{th}$ percentile, it is 8.4%. Since the fraction of stocks with positive $R_{oos}^2$ ranges from 7.7% to 35.8% based on forecasts by the four algorithms, we see the median $R_{oos}^2$ are all negative. Lastly, we report the average $R_{oos}^2$ based on those stock-quarters with positive $R_{oos}^2$. The average $R^2_{oos}$ ranges from 1.6% to 3.6%. The average $R^2_{oos}$ might seem small, but we are making daily return forecasts. Campbell and Thompson (2008) show that at short horizons, small increases in out-of-sample $R^2_{oos}$ can generate large benefits for investors.

*3.2 $R^2_{oos}$, Multinational and Domestic firms, and Industries*

In this subsection, we analyze the stock return predictability of foreign signals for multinational and domestic firms and firms in different industries. We define multinational firms as those with more than 5% of total sales from non-U.S. markets. [13]

In Table 4, we first find that the fraction of stocks with positive $R^2_{oos}$ is almost identical between multinational and domestic firms. For both multinational and domestic firms, we can obtain positive $R^2_{oos}$ for about 56% of their stocks based on return forecasts by at least one of the four algorithms. Among those stocks with positive $R^2_{oos}$ in each quarter, the average $R^2_{oos}$ is almost the same for multinational and domestic firms. These findings unveil a surprising and important message: domestic firms are significantly exposed to what happens in foreign markets, possibly to the same level as multinational firms.

Naturally, we could expect multinational firms to be directly affected by foreign events through their foreign operations. However, the findings on domestic firms highlight machine learning algorithms' ability to identify more complex and indirect channels through which foreign signals impact stock returns. For example, foreign signals could impact U.S. firms' competitors and suppliers in foreign countries, U.S. firms' domestic competitors' operations and suppliers in foreign countries, foreign companies in which U.S. firms have ownership, and so on.

---

[13] Table A2 of the Internet Appendix I reports the fraction of the S&P 500 stocks from multinational and domestic firms and firms in each of the Fama and French 12 industries.

Understanding the transmission channels of these foreign signals for each stock is extremely challenging. But our findings suggest that advanced machine learning algorithms can help.

In addition, we also show the fraction of stocks with positive $R^2_{oos}$ for firms in each of the Fama and French 12 industries. Computers, Software, and Electronic Equipment industry has the highest percentage of firms with positive $R^2_{oos}$, at 60%. Even for the industry with the lowest fraction of stocks with positive $R^2_{oos}$, the Utilities, 49% of the stocks have positive $R^2_{oos}$. The average $R^2_{oos}$ calculated based on stocks with positive $R^2_{oos}$ in each quarter is also very similar across industries. Overall, the results in Table 4 indicate that foreign signals' return predictability is similar across multinational and domestic firms and firms in different industries.

## 4. Signal Importance Analysis

In this section, we unpack the "black boxes" of machine learning models by analyzing signals' importance to model predictions. The signal importance analysis provides evidence that our models capture the impact of foreign information on U.S. stock returns. The results here are based on the return forecasts by the pool of 188 foreign stock market return signals.

We conduct our signal importance analyses based on those stock-quarters with positive out-of-sample $R^2_{oos}$, which suggests the models capture some meaningful relationships between foreign signals and returns. We train models at stock-quarter frequency. The previous section shows that a subset of the S&P 500 stocks is associated with positive out-of-sample $R^2_{oos}$. A negative $R^2_{oos}$ indicates either the corresponding U.S. firm is not exposed to foreign signals, or the model fails to learn the relationship. The observations in our following analyses on signal importance are at the stock-quarter-signal level, with each signal uniquely identified by a country-lagged week combination.

*4.1 Measuring signal importance*

Even though measuring signal importance for tree-based algorithms and neural networks is challenging, the machine learning literature has developed methods to reveal what happened in these black-box models (e.g., Lundberg and Lee (2017), Shrikumar, Greenside, and Kundaje (2019), Lopez de Prado (2020), and Lundberg et al. (2020)). In this study, we measure signal importance using SHAP (SHapley Additive exPlanations), a game theoretic approach to improve model interpretability.[14] This method is built upon the Shapley value concept in game theory (Shapley (1951, 1952)) and introduced by Lundberg and Lee (2017). Finance literature has also adopted it to explain machine learning models (e.g., Erel et al. (2021) and Bali et al. (2023)).

One advantage of SHAP is to understand models at the granular level. For each trained model, we can obtain a signal's SHAP value for each training observation. A signal's SHAP value for one daily observation estimates how much this signal pushes the model's prediction away from the base value (the average prediction over all the training observations). Based on the SHAP values, we can draw graphs to show how the changes in one foreign signal affect one model's return forecasts. In Figure A1 of the Internet Appendix I, we focus on one model trained using Apple's data from 2019Q2 to 2020Q1 based on RF. The horizontal axis shows the lagged 1-week returns for each of the four countries: China, Germany, Mexico, and Saudi Arabia. The vertical axis shows the SHAP values. The four graphs inform us of the model's estimations on the relationships between the lagged 1-week stock market return signals and Apple's daily return forecasts.

For each model, we then aggregate and compute each signal's mean absolute SHAP value based on its SHAP values over all the daily training observations (Lundberg et al. (2020)). A signal's mean absolute SHAP value reflects this signal's average incremental contribution to the model's predictions. The mean absolute SHAP values help us compare the relative importance of different signals. We use the mean absolute SHAP value associated with each model and signal to measure signal importance.

---

[14] We compute SHAP values using the SHAP python library available at https://github.com/slundberg/shap.

Measuring each foreign signal's importance to LASSO's predictions is relatively convenient. Similar to OLS, LASSO estimates coefficients on each independent variable. We use the absolute value of estimated coefficients to quantify the importance of each foreign signal. For LASSO models, as described in subsection 1.3, we standardize our foreign signals to make them have zero mean and a standard deviation of one. The absolute value of the coefficient on a foreign signal reflects the absolute change in predicted returns when that foreign signal changes by one standard deviation.

*4.2 Signal importance in each lagged week and the slow dissemination of foreign information*

In this subsection, we study the relative importance of foreign signals in each of the four lagged weeks. To predict one U.S. stock's return in a day, we use lagged weekly returns of the stock market indices during the previous 4-week period. If the information embedded in these foreign signals is slowly disseminated into U.S. stock prices, signal importance should decrease gradually from those lagged 1-week signals to lagged 4-week signals. Namely, U.S. stock prices have absorbed the information in the lagged 1-week signals to a much lesser extent than in lagged 4-week signals.

In Table 5, we regress signal importance on the lagged week indicator. In Panel A, we run linear regressions. The independent variable, lagged week, takes the value of *i* for lagged *i*-week foreign signals. For example, it is 1 for lagged 1-week foreign signals and 4 for lagged 4-week foreign signals.[15] We find the coefficients on the lagged week indicator are all negative and statistically significant for all four algorithms. These findings confirm that the signal importance decreases from lagged 1-week to lagged 4-week signals.

We also quantify how long it takes for the information in foreign signals to be disseminated into U.S. stock prices by analyzing the magnitude of coefficient estimations and the constant. When we increase the number of lagged weeks, the estimated signal importance from these linear regressions will decrease. The number that makes the estimated signal importance drop to zero or negative would tell us the time window for U.S. stocks to absorb the information in foreign signals fully.

Based on LASSO's estimations, the estimated signal importance turns to negative values for lagged 7-week signals. This finding suggests that our LASSO models estimate that the information in foreign signals takes up to 6 weeks to be fully disseminated into U.S. stock returns. For RF, the estimated signal importance turns to negative values for lagged 19-week signals. For GB, till lagged 38 weeks, the estimated signal importance drops to negative territory. For NN, it is over lagged 100 weeks. It is hard to imagine the information in foreign signals can take multiple years to be incorporated into stock prices. Therefore, our estimated linear relationships between signal importance and the lagged indicator in Panel A may be only qualitatively meaningful. And the importance of foreign signals might decrease non-linearly as the number of lagged weeks increases.

In Panel B, we model the signal importance decreases at an accelerated pace as the number of lagged weeks increases. We still run linear regressions, but the independent variables become the values from the natural exponential function of the lagged week indicator. Figure 3 visualizes the estimated signal importance along each of the lagged weeks based on the estimations in Panel B of Table 5. We still observe a negative relationship between signal importance and the number of lagged weeks. This time, LASSO models estimate it takes up to 4 weeks for information in foreign signals to be fully incorporated into U.S. stock prices. RF models estimate it needs 5 weeks, and GB models estimate it takes 6 weeks. NN models estimate it takes up to 8 weeks for U.S. stocks to absorb such information fully.

---

[15] For the same stock-quarter, the signal importance of the four lagged signals from the same country could be correlated. We compute standard errors clustered by stock × quarter × country to allow for correlations of regression residuals within the same stock, quarter, and country.

LASSO only captures linear relationships between foreign signals and U.S. stock returns. The increased time window to absorb information in foreign signals from LASSO to RF and GB and then to NN implies that tree-based algorithms and neural networks have learned non-linear and more complex links between foreign signals and stock returns.

In Table A3 of the Internet Appendix I, we also run the regressions that include stock fixed effects to control for unobserved stock-level characteristics and quarter fixed effects to control for unobserved variables that change over time. We find very similar coefficients on the lagged week indicators, as in Table 5. In Table 5, we do not include fixed effects. Adding fixed effects would make the magnitude of the intercept hard to interpret. Thus, we cannot use the intercept to estimate how long it takes for the information in foreign signals to be disseminated into U.S. stock prices.

Overall, the findings in this subsection indicate that the information embedded in foreign signals is slowly disseminated into U.S. stock prices. Meanwhile, we admit that the estimated time window for foreign information dissemination here is based on using lagged foreign market returns as foreign signals. The speed of foreign information dissemination could differ when researchers use other types of foreign signals.

*4.3 Factors affecting the speed of foreign information dissemination*

In this subsection, we examine those factors that affect the speed of foreign information dissemination. The factors we consider include media coverage of news regarding a foreign market, whether signals are from developed markets, foreign institutional ownership, multinational vs. domestic firms, and share turnover.

In Table 6, we interact these factors with the lagged week indicator. Negative coefficients on the interaction terms indicate the importance of foreign signals decreases faster as the number of lagged weeks increases; namely, the information in foreign signals is disseminated into U.S. stock prices more quickly.

The visibility of foreign news in the media can accelerate the dissemination of foreign information into stock prices. A large amount of media coverage on news regarding a foreign market can draw investor attention to information in that foreign market, speeding up the price discovery of foreign information.

We count the total number of news stories regarding entities in a foreign market during the four-quarter window for training each model.[16] To mitigate the trend in overall media coverage over time, we scale the total number of news stories regarding a foreign market by the total number of news stories covered by RavenPack during the same period. We define this ratio as the news coverage of a market variable in Table 6. A higher value of this variable indicates higher levels of media coverage regarding a foreign market during the training period for a model. Table 6 shows the coefficients on the interactions between the lagged week indicator and news coverage are negative and statistically significant for RF, GB, and NN. This finding suggests that U.S. stock prices take less time to incorporate information in signals from foreign markets with high media coverage.

---

[16] As in section 2.2, we obtain the news coverage data from RavenPack. RavenPack assigns countries to entities involved in each news story. For any news story that mentions an entity, RavenPack provides a relevance score. We only consider those entities with an above 90 relevance score, indicating those entities are strongly relevant in the news stories and usually referenced in the main title or headline of the news items.

Emerging markets could exhibit poor quality and low credibility of financial information and present high information asymmetries. It would be more efficient for investors to collect and process information in developed markets. Table 6 shows that the coefficients on the interactions between the lagged week indicator and the developed market dummy are negative and statistically significant for RF, GB, and NN, suggesting U.S. stock prices take less time to absorb information in signals from developed markets.

Institutional investors in foreign countries can have advantages in gathering information from their home countries. Prior studies also find that foreign institutional investors improve price informativeness and possess unique information sets relative to domestic investors (e.g., He et al. (2013) and Kacperczyk, Sundaresan, and Wang (2021)). Therefore, we would expect the prices of U.S. stocks with higher foreign institutional ownership to take less time to absorb the information in foreign signals. Our results show that U.S. stocks with higher foreign institutional ownership take less time to absorb the information in foreign signals based on the LASSO, RF, and GB estimations.

Greater overall institutional ownership, in general, improves price informativeness. In Table 6, we do not see a significant impact brought by the overall institutional ownership after we control foreign institutional ownership. These findings indicate that foreign institutional investors bring unique information about foreign markets. And our machine learning models can capture such an effect.

In addition, we use share turnover as the proxy for stock liquidity. The idea is that greater liquidity facilitates the incorporation of information into prices. Bai, Philippon, and Savov (2016) find that stocks with higher share turnover have greater price informativeness. In Table 6, our results also show that U.S. stocks with higher share turnover take much less time to absorb the information in foreign signals.

Lastly, we test whether foreign information dissemination is faster among multinational firms. When a firm generates significant revenues from foreign markets, investors are likely to pay more attention to information relevant to the firm's foreign business. As a result, foreign information would be more efficiently incorporated into multinationals' prices. In Table 6, the estimations based on RF and NN models confirm this conjecture.

On balance, the findings in this subsection provide evidence that our machine learning models can learn the impact of foreign information on U.S. stock prices.

*4.4 The speed of foreign information dissemination over time*

Financial markets have grown substantially in recent decades and become much more liquid. New information technologies have been applied in price discovery, and institutional investors become dominant in the markets. Bai, Philippon, and Savov (2016) show that the price informativeness of the S&P 500 stocks has increased steadily since 1960. In this subsection, we examine whether the price discovery of foreign information has also improved over time. If so, is the improved price discovery more pronounced on information from emerging or developed markets?

In Panel A of Table 7, we first regress signal importance on the lagged week indicator, a later period dummy variable, and the interaction between the lagged week indicator and the later period dummy. The later period dummy takes the value of 1 for signals from models with estimation windows ending in or after 2017. The coefficients on the later period dummy variable are absorbed by the quarter fixed effects.

For LASSO, RF, GB, and NN, the coefficients on the interaction terms are all negative and statistically significant. These results indicate that as the number of lagged weeks increases, the importance of foreign signals decreases faster in the later half of the sample; namely, U.S. stocks take much less time to incorporate foreign information in the later period.

In addition, we introduce the settings involving triple interactions among the developed market dummy, the lagged week indicator, and the later period dummy. For LASSO, RF, and GB, the coefficients on the triple interaction terms are positive and statistically significant. This finding suggests that as time goes by, the improved price discovery of foreign information is more pronounced in emerging-market information. In Panel B of Table 7, we replace the later year dummy variable with a continuous year variable. We find messages similar to those in Panel A.

Given our sample period, we only examine how the speed of foreign information dissemination changes from 2012Q1 to 2022Q1. Our findings suggest that financial markets have become more efficient at incorporating foreign information over this ten-year period. The gains are mainly from the improved price discovery of emerging-market information.

*4.5 COVID-19 crisis and signal importance*

In late December 2019, a previously unknown virus emerged from Wuhan, China. During 2020Q1, it quickly spread from a local public health crisis to a global pandemic. On February 11, 2020, the World Health Organization (WHO) announced the name of the disease that caused the 2019 Novel Coronavirus outbreak: "COVID-19." The COVID-19 pandemic drove millions of people out of work, caused economic shutdowns in many countries, and broke the global supply chain. Equity markets worldwide also plummeted. Austria, Brazil, Columbia, Egypt, Greece, Philippines, and UAE stock markets dropped over 30% in 2020Q1.

The COVID-19 outbreak in foreign markets and the related big swings in foreign stock markets drew investors' attention to foreign information. As a result, investors could trade more using foreign information and speed up the price discovery of the information in foreign signals. In Table 8, we examine whether the outbreak of COVID-19 affects the signal importance in each lagged week. We follow the settings in Table 6 and interact the COVID-19 dummy with the lagged week indicator. COVID-19 is a dummy variable, taking the value of 1 for signals from models with a training period including 2020Q1. For all four algorithms, we find negative coefficients on the interactions, indicating the information in foreign signals was disseminated much faster during the COVID-19 crisis.

*4.6 The five most important foreign markets*

We report the five most important markets over time based on the average signal importance of each market in Table 9. The five most important markets vary significantly for different algorithms. For example, the top five foreign markets identified by LASSO in 2021 are Japan, Saudi Arabia, Thailand, Peru, and UAE. They are New Zealand, UAE, India, Mexico, and Kuwait for RF in 2021. GB highlights Denmark, Norway, Malaysia, Kuwait, and Peru as the five most important markets in 2021. For NN in 2021, they are Belgium, Colombia, Kuwait, Indonesia, and Chile. For the same algorithm, we also see the importance of each market changes over time. For instance, the five most important markets are Qatar, Turkey, Thailand, Denmark, and Mexico for GB in 2012. They change to China, Peru, Kuwait, New Zealand, and Russia in 2016.

More importantly, the list of the most important foreign markets also reveals the potential difference between human intuition and machine learning algorithms. Very likely, an ordinary investor would conjecture the most important markets are those with the largest gross domestic product (GDP) or stock market capitalization. However, Table 9 shows this is not the case. For example, Qatar frequently emerges as one of the most important foreign markets. Perhaps upon reviewing the list in Table 9, this investor grasps Qatar's critical role in the global supply of natural gas and oil. Additionally, Qatar often acts as a key mediator in those geopolitical crises of the Middle East.[17]

*4.7 Economic links with the U.S. market*

In this subsection, we examine whether signals from foreign markets with stronger economic links with the U.S. carry greater importance. We consider four factors reflecting the economic connections between a foreign market and the U.S.: international trade, geopolitical risk, economic uncertainty, and foreign sales. International trade is the U.S. trade in goods with a foreign market in $trillion. A foreign market's international trade with the U.S. directly reflects the business connections between the U.S. and the foreign market. Geopolitical risk is based on the geopolitical risk index constructed by Caldara and Iacoviello (2022), who find that high geopolitical risk leads to a decline in real economic activity. Economic uncertainty is based on the world uncertainty index constructed by Ahir, Bloom, and Furceri (2022), who show that the index is negatively associated with GDP growth and output. It is challenging for U.S. firms to build business relationships in countries with political upheaval, war threats, or shaky economic and financial systems. Thus, signals from markets with higher geopolitical risk and economic uncertainty should be less important to models' predictions. Foreign sales is a dummy variable, taking the value of 1 if a U.S. firm has at least 5% of its total sales from a foreign market. When a U.S. multinational directly generates revenues from a foreign market, we may expect signals from that market to matter.

---

[17] For example, see https://www.cnn.com/2023/11/01/middleeast/qatar-mediation-israel-hamas-intl/index.html. https://foreignpolicy.com/2023/10/28/qatar-middle-east-israel-hostages-hamas-gaza-mediator/.

In Table A4 of the Internet Appendix I, we regress signal importance on international trade, geopolitical risk, economic uncertainty, and foreign sale dummy. International trade is positively and significantly related to signal importance for models by LASSO, RF, and GB. Geopolitical risk and economic uncertainty are negatively associated with signal importance for all four algorithms' models. For RF, we find that the foreign sale dummy is positively and statistically significantly related to signal importance. In sum, these findings provide another piece of evidence that our machine learning models reflect the impact of foreign information on U.S. stock returns. Signals from foreign markets with more international trade with the U.S. and stabler business environments are more important to models' return forecasts.

*4.8 Stock-specific time-varying relationships*

This subsection analyzes signal importance to show the relationships between foreign signals and U.S. stock returns are stock-specific and time-varying. Our estimated signal importance is at the *stock-country-lagged week-quarter* level, and a combination of *country-lagged week* uniquely identifies one foreign signal. In Panel A, Table A5 of the Internet Appendix I, we first regress signal importance on *country × lagged week × quarter* fixed effects. If the impact of one foreign signal on different stocks' returns is the same in a quarter, the signal importance variation within fixed effects would be zero. The *country × lagged week × quarter* fixed effects should explain 100% variation of signal importance. However, we find the $R^2$ is below 15% for LASSO, RF, and GB and is 74% for NN. These results suggest that the same foreign signal has a distinct impact on different U.S. stocks' returns.

In Panel B of Table A5, we regress signal importance on *stock × country × lagged week* effects. If the relationship between one foreign signal and one stock's returns is constant over time, the signal importance variation within fixed effects would be zero. And the *stock × country ×* *lagged week* fixed effects should explain 100% variations of signal importance. But the $R^2$ is below 30% for LASSO, RF, and GB and is 31% for NN. These findings indicate that the relationship between one foreign signal and one U.S. stock's returns changes substantially over time.

## 5. Predicting the S&P 500 Index Returns

In this section, we examine whether foreign signals can predict the market returns, even though they exhibit return predictability for a subset of individual U.S. stocks. We build bottom-up forecasts for the S&P 500 index daily returns by aggregating and value-weighting individual stock return forecasts.

We first examine the performance of market-timing strategies based on these predicted daily returns for the S&P 500 index. The market-timing strategies follow a simple rule. On the upside, when the bottom-up forecasts by all four algorithms (LASSO, RF, GB, and NN) are positive for a day, we choose 200% or 300% exposure to the S&P 500 index. On the downside, we give -100% exposure to the S&P 500 index when all four bottom-up forecasts are negative for a day. When the forecasts by all four algorithms do not have a consensus, we choose 100% exposure to the S&P 500 index.[18]

We present the performance of market-timing strategies in Panel A of Table A6 of the internet appendix. Column (1) shows the average return of the market-timing strategy is 7.66 basis points per day with a t-statistic of 2.78 (or 19.3% annually), when we choose 200% exposure on the upside. In contrast, Column (3) shows the average daily return of the S&P 500 index is 5.30 basis points during the same period. Column (4) shows the statistically significant difference in average daily returns between our market-timing strategy and the S&P 500 index. Meanwhile, our market-timing strategy has a higher Sharpe ratio and a slightly lower maximum one-quarter loss.

---

[18] Investors can conveniently achieve leveraged daily exposure to the S&P 500 index through leveraged exchange-traded funds (ETFs). For example, ProShares Ultra S&P500 ETF, launched in 2006, provides 200% daily exposure to the S&P 500 index with a 0.89% annual expense ratio. ProShares UltraPro S&P500 ETF, launched in 2009, provides 300% daily exposure to the S&P 500 index with a 0.91% annual expense ratio. Negative exposure can also be accessible through leveraged ETFs. For example, ProShares Short S&P500 ETF, launched in 2006, provides -100% daily exposure to the S&P 500 index with a 0.89% annual expense ratio. To get 100% exposure to the S&P 500 index, investors can easily find mutual funds and ETFs tracking the performance of the S&P 500 at minimal expense ratios.

Column (2) reports that the average return of the market-timing strategy increases to 10.94 basis points per day with a t-statistic of 3.00, when we use 300% exposure on the upside. This time, the Sharpe ratio of the market-timing strategy also increases to 0.94, but its maximum one-quarter loss is more than the S&P 500 index. We also compute the portfolio turnover. The turnover is 100% in a day if the strategy changes the exposure to the S&P 500 index from the previous day and 0% otherwise. We find that the average daily turnover is 28.5%. In addition, Figure A2 of the Internet Appendix I presents the cumulative returns of the market-timing strategies.

In Panel B of Table A6, we compute the out-of-sample predictive $R^2_{oos}$ based on our bottom-up daily forecasts and real daily returns of the S&P 500 index for each quarter. Column (1) shows the fraction of quarters in our sample associated with positive $R^2_{oos}$. The bottom-up forecasts based on LASSO exhibit positive predictability in 56% of the quarters. The bottom-up RF forecasts show positive predictability in 39% of the quarters. For GB, the number is 49%. For NN, we do not see any quarter associated with positive $R^2_{oos}$. This is due to the forecasts by NN giving positive $R^2_{oos}$ for only 7.7% of the S&P 500 stocks as shown in Table 3. When we build the bottom-up forecasts by aggregating all the predicted returns of S&P 500 stocks by NN, we include too many stocks' noisy forecasts. Overall, 68% of the quarters are associated with positive $R^2_{oos}$ based on bottom-up forecasts by at least one of the LASSO, RF, or GB algorithms.

Columns (2) to (5) report the statistics of $R^2_{oos}$. The maximum $R^2_{oos}$ reported in Column (4) suggests that in some quarters, the bottom-up return forecasts can explain more than 5% of the variations of the daily S&P 500 index returns. Column (5) reports the average $R^2_{oos}$ based on those quarters associated with positive $R_{oos}^2$. For LASSO, the average $R_{oos}^2$ is 2.09%; for RF, the average $R_{oos}^2$ is 3.21%; for GB, the average $R_{oos}^2$ is 2.54%.

Previous studies show that lagged non-U.S. stock market returns could not predict the U.S. market returns, while lagged U.S. market returns significantly predict market returns in non-U.S. countries (Rapach, Strauss, and Zhou (2013)). Our results here suggest that the bottom-up forecasts aggregating individual stock return predictions using lagged non-U.S. market returns exhibit out-of-sample predictability for the U.S. stock market returns.

**6. Searching A Pool of ≈13,000 Candidate Foreign Signals**

In this study, we further challenge machine learning algorithms with another significantly larger pool of candidate foreign signals. This pool includes lagged weekly returns of individual stocks and stock markets in 47 non-U.S. markets during the previous four weeks. Individual stock return signals are based on all the stocks with a top 10% market capitalization in each country. This approach results in a pool of around 13,000 candidate foreign signals. For computational reasons, we only use LASSO to search this large pool. We follow the training and forecasting procedures described in previous sections.[19]

*6.1 Portfolio performance*

In Table 10, we report out-of-sample $R_{oos}^2$. When we compute $R_{oos}^2$ using predictions by this pool of ≈13,000 signals (LASSO_13,000), on average, about 12% of the S&P 500 stocks have positive out-of-sample predictive $R_{oos}^2$ in a quarter. And the average $R_{oos}^2$ based on those stock-quarters with positive $R_{oos}^2$ is 1.6%. These findings suggest that LASSO can identify valuable signals from this large pool of ≈13,000 signals to forecast daily U.S. stock returns.

---

[19] OLS regressions cannot be applied to this setting in the first place, because the number of candidate signals way exceeds the number of observations in our training sample.

In Table 3, we find that 23.5% of the S&P 500 stocks have positive $R^2_{oos}$ in a quarter, when we use the forecasts by LASSO and the 188 stock market return signals (LASSO_188). Panel A of Table A7 of the Internet Appendix I shows the fraction increases to 29.0%, when we consider the stocks with the positive $R^2_{oos}$ based on forecasts by either LASSO_13,000 or LASSO_188. In Table 3, 56.2% of the S&P 500 stocks have positive $R^2_{oos}$ in a quarter, when we consider the forecasts by LASSO, RF, GB, or NN. With the addition of LASSO_13,000, Panel A of Table A7 shows the fraction increases to 59.0%. These results suggest adding individual foreign stock returns as candidate signals provides incremental predictive power.

In Panel B, Table A7 of the Internet Appendix I, we present the performance of portfolios formed on return forecasts by this large pool of foreign signals. In Column (1), we construct a portfolio of stocks with top 5% forecasted returns by LASSO and equally weight each stock. The average return of this portfolio in excess of the daily one-month U.S. Treasury yield is 9.52 basis points per day with a t-statistic of 3.07. The six-factor alpha of this portfolio is 2.68 basis points per day with a t-statistic of 1.93. Column (2) shows the six-factor alpha of the portfolio consisting of stocks with bottom 5% forecasted returns is -2.44 basis points per day with a t-statistic of -1.77. In Column (3), we construct a long-short portfolio that buys stocks with top 5% forecasted returns and sells short stocks with top 5% forecasted returns. The six-factor alpha of this long-short portfolio is 5.12 basis points per day with a t-statistic of 2.79.

*6.2 Signal importance analysis*

Using individual foreign stock return signals provides an additional dimension to analyze signal importance. When we use human intuition to identify foreign signals, it seems natural to prioritize or even focus on foreign firms in the same industry as U.S. firms. Panel C of Table A7 assesses whether signals from foreign firms in the same industry as U.S. firms carry greater importance. For consumer nondurables, and telephone and television transmission, the signals from foreign firms in the same industry are more important than other foreign signals. We find the opposite for chemicals and computers, software, and electronic equipment. The relationship is insignificant among other industries. Importantly, when considering all industries collectively, we find that valuable foreign signals are not concentrated in foreign firms in the same industries as U.S. firms.

In Panel D of Table A7, we confirm that the information in foreign signals is slowly disseminated into U.S. stock prices. And it takes up to 4 weeks for information in foreign signals to be fully incorporated into U.S. stock prices. The speed is also much slower for signals from low-media-coverage markets and among stocks with lower foreign institutional ownership.

*6.3 Why does a pool of ≈13,000 candidate foreign signals not lead to better forecasts?*

The findings in the previous subsection suggest LASSO can identify meaningful foreign signals from the large pool of ≈13,000 candidate signals. However, the accuracy of return forecasts by this large pool is not significantly better than those using only 188 foreign stock market return signals. For example, 23.5% of the S&P 500 stocks have positive $R^2_{oos}$ in a quarter when we use the forecasts by LASSO and only stock market return signals. But only 12% of the S&P 500 stocks have positive $R^2_{oos}$ in a quarter when we compute $R^2_{oos}$ using predictions by the large pool of ≈13,000 candidate foreign signals.

When adding individual foreign stocks' returns as candidate signals, we increase the chance of uncovering meaningful relationships between foreign signals and U.S. stock returns. But, at the same time, we add substantial noisy signals and increase the sparsity of valuable signals in the pool. Naturally, we would expect only a small set of foreign firms to have significant business connections with a particular U.S. firm. Consequently, the heightened sparsity also increases the difficulty for machine learning algorithms to detect valuable signals.

Panel E of Table A7 reports the fraction of candidate foreign signals with non-zero coefficients, namely those signals identified by LASSO as valuable predictors. For the pool of 188 foreign market return signals, LASSO selects 2.24% of them to forecast returns. Out of the ≈13,000 candidate signals, the substantial noises in the pool and the increased difficulties in detecting valuable signals due to the noises make only 0.054% of candidate foreign signals selected by LASSO. On balance, our results suggest that the increased sparsity of valuable foreign signals in the pool could hinder the performance of machine learning algorithms. Further expanding the pool of candidate signals — such as incorporating the lagged returns of all individual foreign stocks — might not necessarily improve the performance of return forecasts.

## 7. Robustness Tests

In this section, we report a series of robustness checks. In Table A8 of the Internet Appendix I, we conduct the subperiod analysis for Panel A of Table 1. We find that portfolios formed on return forecasts by foreign signals deliver superior returns pre- and post-2017. Table A9 of the Internet Appendix I indicates that portfolios formed on forecasts by foreign signals generate abnormal returns after controlling for global ex-U.S. risk factors (Fama and French (2016)). In Table A10 of the Internet Appendix I, we aggregate daily portfolio returns to monthly returns. Our results confirm that the findings using daily returns hold at the monthly level.

The objective function in the training process is the mean squared prediction errors, which equally weight the forecast errors of each stock. Therefore, it's natural to study equal-weighted portfolios in our analysis. Our sample focuses on the S&P 500 stocks, so illiquidity and trading costs are not significant concerns. Moreover, using the S&P 500 stocks makes the portfolio of stocks with top (bottom) 5% forecasted returns include around 25 stocks. Thus, the value-weighted portfolio has the disadvantage of being dominated by a few stocks with substantially large market capitalization. Table A11 of the Internet Appendix I uses value-weighted portfolios to examine the daily performance of portfolios formed on returns forecasts by foreign signals. We find that the six-factor alpha of the long-short portfolios based on forecasts by LASSO, RF, and GB is not statistically significant. The six-factor alpha of the long-short portfolios based on forecasts by NN and the combined portfolio is statistically significant.

Zou and Hastie (2005) argue that LASSO tends to select only one predictor from a group of highly correlated predictors and design the Elastic Net algorithm to encourage a grouping effect, where strongly correlated predictors tend to be in or out of the model together. In Table A12 of the Internet Appendix I, we follow our training and predicting procedures and use Elastic Net to forecast returns. We present portfolio performance and out-of-sample $R^2_{oos}$ using the forecasts generated by Elastic Net. The results are very similar to the ones based on LASSO's return forecasts.

## 8. Conclusion

This paper introduces a new machine learning approach to detect value-relevant foreign information for both domestic and multinational companies. Rather than solely using human reasoning, we take a different approach and completely rely on machine learning algorithms to detect valuable foreign signals. We train over 100,000 models to capture stock-specific time-varying relationships between foreign signals and returns. We find that foreign signals exhibit out-of-sample return predictability for a subset of U.S. stocks, spanning across domestic and multinational companies. Signal importance analysis reveals that the price discovery of foreign information is much slower for information from emerging and low-media-coverage markets.

Firms use information encoded in asset prices when making real investment decisions. The delays in incorporating foreign information could translate into real investment distortions and misallocations of resources across firms. As more investors adopt machine learning-based investment strategies leveraging foreign signals, we collectively improve the price discovery of foreign information.

While this paper contributes valuable insights into harnessing machine learning's ability to identify foreign signals, we focus on foreign stock market returns and individual foreign stock returns as candidate foreign signals. Machine learning algorithms have also shown their ability to process images and natural languages. Future research can explore a wider array of foreign signals, such as events and announcements documented in textual and visual data from various sources. By incorporating a more diverse range of foreign signals, researchers could uncover a broader and richer impact of foreign information on asset prices. Such analysis would also provide insights into developing disclosure practices and market mechanisms to improve the market efficiency of incorporating foreign information.

Huang, Xing. 2015. "Thinking Outside the Borders: Investors' Underreaction to Foreign Operations Information." *The Review of Financial Studies* 28 (11): 3109–52.

Hsieh, Tien-Shih, Jeong-Bon Kim, Ray R. Wang, and Zhihong Wang. 2020. "Seeing Is Believing? Executives' Facial Trustworthiness, Auditor Tenure, and Audit Fees." *Journal of Accounting and Economics* 69 (1): 101260.

Kacperczyk, Marcin, Savitar Sundaresan, and Tianyu Wang. 2020. "Do Foreign Institutional Investors Improve Price Efficiency?" *The Review of Financial Studies*, July.

Kho, Bong-Chan, René M. Stulz, and Francis E. Warnock. 2009. "Financial Globalization, Governance, and the Evolution of the Home Bias." *Journal of Accounting Research* 47 (2): 597–635.

Khurana, Inder K., Raynolde Pereira, and K. K. Raman. 2003. "Does Analyst Behavior Explain Market Mispricing of Foreign Earnings for U.S. Multinational Firms?" *Journal of Accounting, Auditing & Finance* 18 (4): 453–77.

Langberg, Nisan, and K. Sivaramakrishnan. 2010. "Voluntary Disclosures and Analyst Feedback." *Journal of Accounting Research* 48 (3): 603–46.

Leippold, Markus, Qian Wang, and Wenyu Zhou. 2022. "Machine Learning in the Chinese Stock Market." *Journal of Financial Economics* 145 (2, Part A): 64–82.

Lewellen, Jonathan. 2015. "The Cross-Section of Expected Stock Returns." *Critical Finance Review* 4 (1): 1–44.

Li, Ningzhong, Scott Richardson, and İrem Tuna. 2014. "Macro to Micro: Country Exposures, Firm Fundamentals and Stock Returns." *Journal of Accounting and Economics* 58 (1): 1–20.

Li, Kai, Feng Mai, Rui Shen, and Xinyan Yan, 2021, Measuring Corporate Culture Using Machine Learning, *The Review of Financial Studies* 34, 3265–3315.

Lopez de Prado, Marcos. 2020. "Interpretable Machine Learning: Shapley Values." SSRN Scholarly Paper. Rochester, NY.

Luo, Yuanzhi. 2005. "Do Insiders Learn from Outsiders? Evidence from Mergers and Acquisitions." *The Journal of Finance* 60 (4): 1951–82.

Lundberg, Scott, and Su-In Lee. 2017. "A Unified Approach to Interpreting Model Predictions." arXiv.

**Figure 1: The Cumulative Returns of Lockheed Martin, McDonald's and Starbucks, around the Outbreak of the 2022 Russian-Ukraine War**

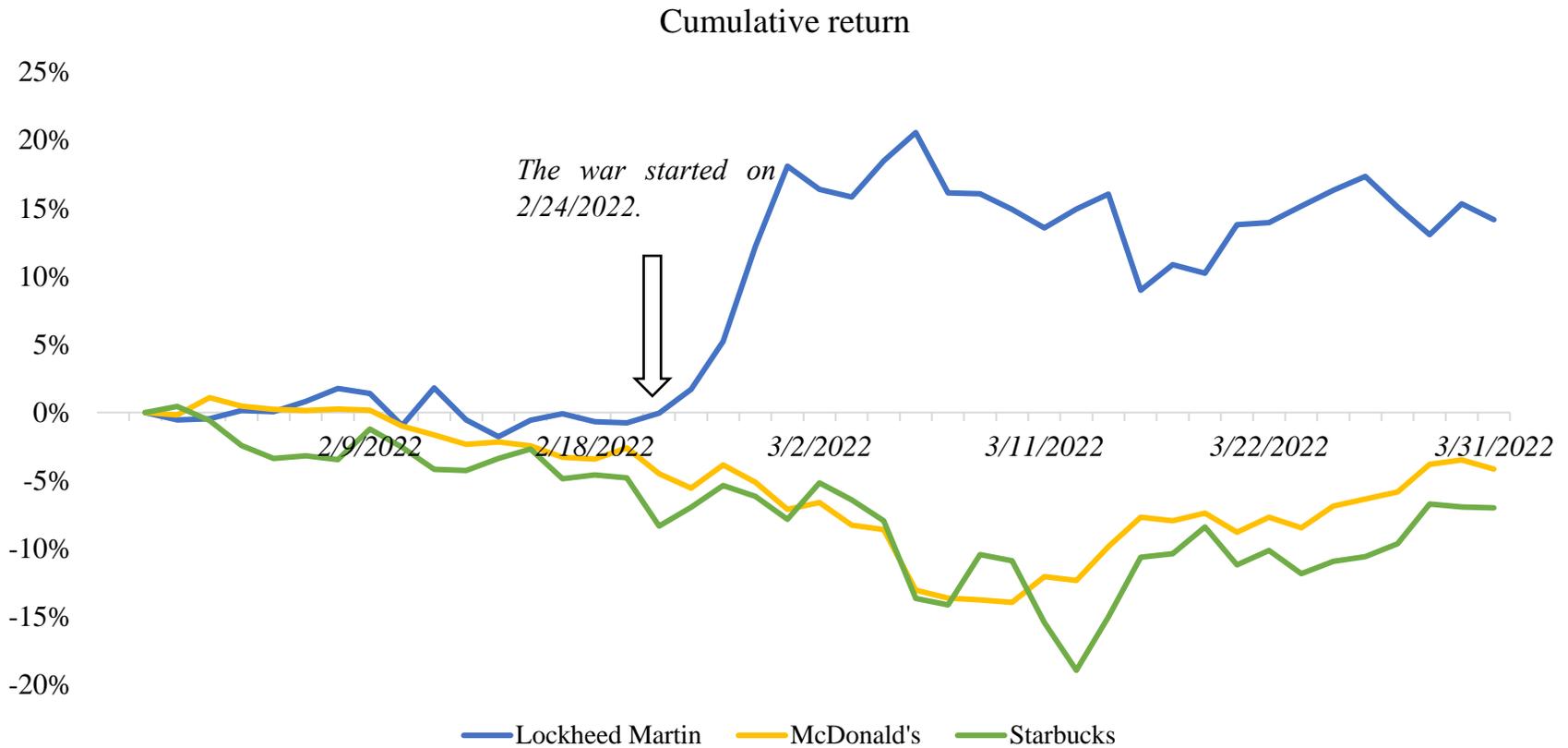

This figure shows the stock performance of Lockheed Martin, McDonald's, and Starbucks in 02/2022 and 03/2022, around the Outbreak of the 2022 Russian-Ukraine War.

**Figure 2: Model Training and Predicting Procedures**

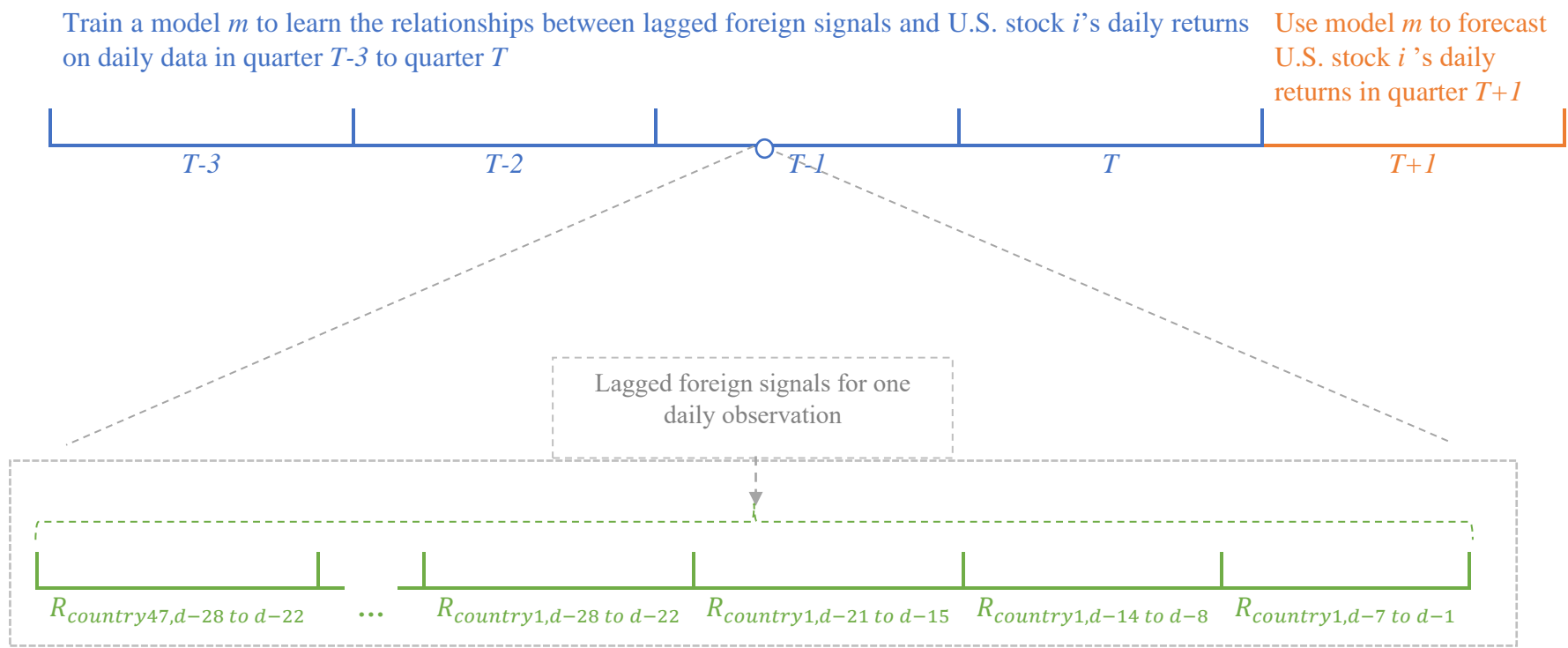

This figure shows one model's training and predicting procedures. We train a model $m$ using one machine learning algorithm based on daily data from quarter $T$-$3$ to quarter $T$. As illustrated in the box surrounded by dotted lines, for each daily observation, we link U.S. stock $i$'s return on that day with lagged weekly stock market returns in 47 non-U.S. markets during the previous four weeks ($R_{country1,d-7\ to\ d-1}$, $R_{country1,d-14\ to\ d-8}$, ..., $R_{country47,d-28\ to\ d-22}$). To forecast U.S. stock $i$'s returns on each trading day in quarter $T$+$1$, we feed the trained model $m$ with corresponding lagged weekly foreign market returns. When repeating the steps for all the quarters in the sample, we obtain U.S. stock $i$'s return forecasts for all the trading days. We then implement the whole procedure for each stock in our sample to generate return forecasts for all the stocks.

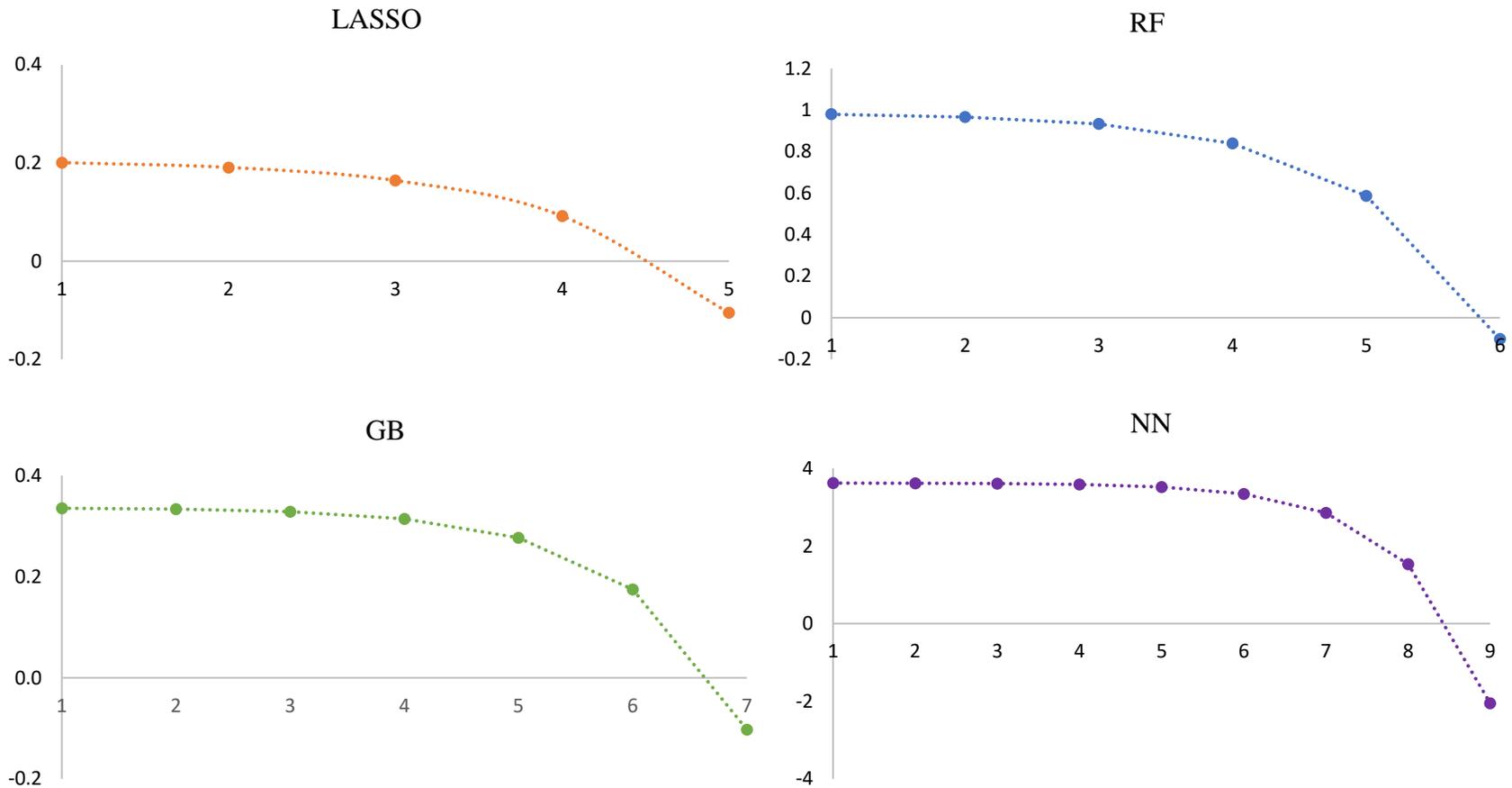

Figure 3: Signal Importance in Each Lagged Week

This figure presents the relationships between the lagged week indicator and signal importance. The horizontal axis shows the lagged week indicator, which takes the value of $i$ for lagged $i$-week foreign signals. For example, it is 1 for lagged 1-week foreign signals and 4 for lagged 4-week foreign signals. The vertical axis shows the signal importance. For LASOO, the signal importance is the absolute value of the coefficient. The signal importance is the SHAP value for RF, GB, and NN. We multiply the original signal importance values by $10^{\wedge}4$. Figures are drawn based on the estimations in Panel B of Table 5.

**Table 1: Performance of Portfolios Formed on Return Forecasts by Foreign Signals**

Panel A: Daily Portfolio Returns

|  | LASSO | | | RF | | | GB | | |
|---|---|---|---|---|---|---|---|---|---|
|  | (1) | (2) | (3) | (4) | (5) | (6) | (7) | (8) | (9) |
|  | Top | Bottom | T-B | Top | Bottom | T-B | Top | Bottom | T-B |
| Mean return | 10.40*** | 4.83 | 5.57** | 9.66*** | 5.88** | 3.78** | 9.72*** | 5.87* | 3.86* |
|  | (3.44) | (1.41) | (2.43) | (3.57) | (2.02) | (2.36) | (3.78) | (1.78) | (1.68) |
| Alpha | 3.69*** | -1.94 | 5.63*** | 3.22*** | -0.59 | 3.81** | 3.23*** | -0.57 | 3.80** |
|  | (2.64) | (-1.36) | (2.77) | (3.05) | (-0.51) | (2.51) | (3.14) | (-0.44) | (2.34) |
| MKT | 1.16*** | 1.21*** | -0.05 | 1.11*** | 1.12*** | -0.01 | 1.09*** | 1.15*** | -0.07 |
|  | (55.68) | (28.11) | (-1.08) | (65.30) | (41.66) | (-0.46) | (82.04) | (28.73) | (-1.59) |
| SMB | 0.37*** | 0.31*** | 0.06 | 0.23*** | 0.27*** | -0.04 | 0.18*** | 0.32*** | -0.14*** |
|  | (10.31) | (7.83) | (1.23) | (9.89) | (9.85) | (-1.08) | (7.49) | (8.44) | (-2.94) |
| HML | 0.32*** | 0.19*** | 0.13** | 0.23*** | 0.19*** | 0.04 | 0.25*** | 0.24*** | 0.00 |
|  | (7.35) | (4.63) | (2.19) | (7.49) | (6.19) | (0.98) | (8.93) | (6.49) | (0.05) |
| MOM | -0.07*** | -0.64*** | 0.57*** | -0.13*** | -0.38*** | 0.25*** | 0.15*** | -0.63*** | 0.78*** |
|  | (-2.82) | (-19.43) | (12.37) | (-3.69) | (-17.27) | (6.97) | (8.13) | (-25.92) | (26.25) |
| RMW | -0.04 | -0.16*** | 0.12 | -0.06* | -0.02 | -0.04 | -0.10*** | -0.03 | -0.06 |
|  | (-0.70) | (-3.18) | (1.63) | (-1.67) | (-0.54) | (-0.68) | (-2.80) | (-0.74) | (-1.12) |
| CMA | 0.01 | 0.20*** | -0.19* | -0.01 | 0.17*** | -0.18** | 0.07 | 0.12* | -0.04 |
|  | (0.11) | (3.02) | (-1.87) | (-0.13) | (3.35) | (-2.57) | (1.39) | (1.72) | (-0.50) |
| Observations | 2,579 | 2,579 | 2,579 | 2,579 | 2,579 | 2,579 | 2,579 | 2,579 | 2,579 |
| Adjusted $R^2$ | 0.78 | 0.82 | 0.19 | 0.85 | 0.84 | 0.10 | 0.84 | 0.84 | 0.47 |

|  | NN | | | Comb | | | OLS | | |
|---|---|---|---|---|---|---|---|---|---|
|  | (10) | (11) | (12) | (13) | (14) | (15) | (16) | (17) | (18) |
|  | Top | Bottom | T-B | Top | Bottom | T-B | Top | Bottom | T-B |
| Mean return | 9.70*** | 3.68 | 6.02*** | 9.87*** | 5.06 | 4.81*** | 6.07** | 7.33*** | -1.25 |
|  | (3.78) | (1.18) | (2.69) | (3.82) | (1.64) | (2.91) | (2.08) | (2.61) | (-0.58) |
| Alpha | 3.28*** | -2.49* | 5.77*** | 3.36*** | -1.40 | 4.75*** | -0.53 | 1.38 | -1.92 |
|  | (2.77) | (-1.79) | (3.09) | (3.87) | (-1.34) | (3.80) | (-0.39) | (1.05) | (-0.90) |
| MKT | 1.07*** | 1.11*** | -0.04 | 1.10*** | 1.15*** | -0.04 | 1.14*** | 1.04*** | 0.09** |
|  | (43.18) | (50.63) | (-1.31) | (81.60) | (39.85) | (-1.56) | (32.70) | (41.92) | (2.53) |
| SMB | 0.15*** | 0.24*** | -0.09* | 0.23*** | 0.29*** | -0.05 | 0.28*** | 0.22*** | 0.06 |
|  | (4.06) | (6.83) | (-1.71) | (10.94) | (9.77) | (-1.55) | (6.55) | (6.99) | (1.06) |
| HML | 0.15*** | 0.14*** | 0.01 | 0.23*** | 0.19*** | 0.04 | 0.22*** | 0.20*** | 0.02 |
|  | (4.36) | (3.70) | (0.19) | (9.25) | (6.44) | (1.33) | (5.08) | (5.70) | (0.37) |
| MOM | 0.04 | -0.59*** | 0.63*** | -0.00 | -0.56*** | 0.56*** | -0.14** | -0.35*** | 0.22* |
|  | (0.89) | (-13.81) | (15.80) | (-0.18) | (-21.57) | (22.53) | (-2.55) | (-5.48) | (1.88) |
| RMW | 0.02 | -0.09** | 0.11* | -0.04 | -0.08** | 0.03 | -0.03 | -0.11** | 0.08 |
|  | (0.50) | (-2.00) | (1.75) | (-1.47) | (-2.06) | (0.79) | (-0.72) | (-2.27) | (1.10) |
| CMA | 0.04 | 0.06 | -0.01 | 0.03 | 0.14*** | -0.11* | -0.01 | 0.11 | -0.11 |
|  | (0.83) | (0.79) | (-0.17) | (0.65) | (2.61) | (-1.78) | (-0.09) | (1.63) | (-1.05) |
| Observations | 2,579 | 2,579 | 2,579 | 2,579 | 2,579 | 2,579 | 2,579 | 2,579 | 2,579 |
| Adjusted $R^2$ | 0.79 | 0.81 | 0.30 | 0.89 | 0.88 | 0.41 | 0.78 | 0.79 | 0.04 |

Panel B: Foreign News Coverage and Six-factor Alpha

| | Top | | Bottom | | T-B | |
|---|---|---|---|---|---|---|
| | \multicolumn{6}{c}{*Media Coverage of Foreign News*} | | | | | |
| | (1) | (2) | (3) | (4) | (5) | (6) |
| | *High* | *Low* | *High* | *Low* | *High* | *Low* |
| LASSO | 1.29 | 5.54*** | 0.70 | -4.69** | 0.59 | 10.22*** |
| | (0.72) | (2.60) | (0.41) | (-2.05) | (0.24) | (3.16) |
| RF | 1.86 | 4.03** | 0.63 | -1.98 | 1.23 | 6.01*** |
| | (1.36) | (2.54) | (0.43) | (-1.12) | (0.62) | (2.60) |
| GB | 1.97 | 4.11*** | 0.75 | -2.15 | 1.21 | 6.26** |
| | (1.42) | (2.69) | (0.47) | (-1.06) | (0.57) | (2.51) |
| NN | 3.20** | 3.43* | -1.32 | -3.85* | 4.52* | 7.28*** |
| | (2.05) | (1.91) | (-0.75) | (-1.84) | (1.84) | (2.62) |
| Comb | 2.08* | 4.28*** | 0.19 | -3.17* | 1.89 | 7.44*** |
| | (1.82) | (3.30) | (0.15) | (-1.93) | (1.19) | (3.87) |

This table presents the daily performance of portfolios formed on return forecasts by foreign signals. To predict the daily returns of the S&P 500 stocks, we use foreign signals and various machine learning algorithms, least absolute shrinkage and selection operator (LASSO) regression, random forest (RF), gradient boosted tree (GB), neural network (NN), and ordinary least squares regression (OLS). Foreign signals are lagged weekly returns of leading stock market indices in 47 non-U.S. markets during the previous four weeks. In Panel A, *Top* refers to the portfolios of stocks with top 5% return forecasts. *Bottom* means the portfolios of stocks with bottom 5% return forecasts. We equally weight stocks in each portfolio. *T-B* is the portfolio that buys stocks with top 5% return forecasts and sells short stocks with bottom 5% return forecasts. Portfolios are rebalanced every day. We report the mean return and intercept (Alpha) and loadings from regressions of Top or Bottom portfolio returns in excess of the daily one-month U.S. Treasury yield on the Fama and French (1993, 2015) MKT, SMB, HML, RMW, CMA, and Carhart (1997) MOM factors. We also report the mean return and intercept (Alpha) and loadings from the regression of T-B portfolio returns on the same set of factors. In columns (13) to (15), we equally weight the returns of portfolios based on LASSO, RF, GB, and NN for Top, Bottom, or T-B, respectively. Panel B separates all the trading days into those with high or low media coverage of foreign news. Media coverage of foreign news is the number of news stories regarding entities in all 47 foreign markets during the four weeks before a trading day, scaled by the total number of news stories covered by RavenPack in the same period. High (low) media coverage of foreign news indicates those trading days with above (below)-median foreign news coverage during the previous four weeks. We report the six-factor alphas in Panel B. All results are in basis points. Robust *t*-statistics are reported in parentheses. *, **, and *** denote statistical significance at the 10%, 5%, and 1% level, respectively.

Table 2: Performance of Portfolios Formed on Return Forecasts by Foreign Signals, Each Decile

|  | LASSO | RF | GB | NN | Comb | OLS |
|---|---|---|---|---|---|---|
|  | (1) | (2) | (3) | (4) | (5) | (6) |
| High (10) | 1.41 | 1.80** | 2.25*** | 2.13** | 1.90*** | -0.32 |
| 9 | -0.32 | -0.08 | 0.13 | -0.02 | -0.07 | -0.17 |
| 8 | -0.92 | 0.29 | -0.87 | -0.69 | -0.55 | -0.07 |
| 7 | -1.25** | 0.48 | -0.62 | 0.33 | -0.27 | -0.06 |
| 6 | 0.02 | 0.56 | 0.67 | 0.37 | 0.40 | -0.23 |
| 5 | 0.94* | -0.73 | 0.03 | -0.19 | 0.01 | 0.08 |
| 4 | 0.49 | -0.05 | -0.37 | 0.06 | 0.03 | -0.13 |
| 3 | 0.33 | -0.15 | -0.30 | 0.14 | 0.00 | -0.20 |
| 2 | 1.05 | -0.50 | 0.09 | 0.40 | 0.26 | 0.65 |
| Low (1) | -1.05 | -0.90 | -0.27 | -1.81* | -1.01 | 1.24 |
| High - Low | 2.46 | 2.70** | 2.52** | 3.95*** | 2.91*** | -1.56 |

This table presents the daily performance of portfolios formed on return forecasts by foreign signals. To predict the daily returns of the S&P 500 stocks, we use foreign signals and various machine learning algorithms, least absolute shrinkage and selection operator (LASSO) regression, random forest (RF), gradient boosted tree (GB), neural network (NN), and ordinary least squares regression (OLS). Foreign signals are lagged weekly returns of leading stock market indices in 47 non-U.S. markets during the previous four weeks. We sort the S&P 500 stocks into deciles based on each model's return forecasts. We equally weight stocks in each decile. Portfolios are rebalanced every day. In column (5), we equally weight the returns of portfolios based on LASSO, RF, GB, and NN in each decile. We report the intercept (Alpha) from regressions of daily returns of each decile portfolio in excess of daily one-month U.S. Treasury yield on the Fama and French (1993, 2015) MKT, SMB, HML, RMW, CMA, and Carhart (1997) MOM factors. We also report the intercept (Alpha) from the regression of daily returns of portfolios that buy stocks in decile 10 and sell short stocks in decile 1 on the same set of factors. *, **, and *** denote statistical significance at the 10%, 5%, and 1% level, respectively.

# Table 3: Out-of-sample $R^2_{oos}$

| | Fraction of Stocks with $R^2_{oos}>0$ | $R^2_{oos}$ Distribution | | | | | | | | Mean (Stocks with $R^2_{oos}>0$) |
|---|---|---|---|---|---|---|---|---|---|---|
| | | 99th | 95th | 90th | 75th | 50th | 25th | 10th | 5th | |
| LASSO | 23.5% | 5.0% | 2.5% | 1.4% | 0.0% | -0.3% | -2.0% | -5.1% | -8.7% | 1.6% |
| RF | 21.6% | 8.4% | 4.5% | 2.6% | -0.6% | -4.5% | -9.2% | -14.7% | -18.7% | 3.0% |
| GB | 35.8% | 6.3% | 3.8% | 2.6% | 0.8% | -1.0% | -3.1% | -5.4% | -7.1% | 2.0% |
| NN | 7.7% | 7.0% | 1.7% | -1.1% | -7.5% | -21.4% | -53.7% | -167.5% | -400.0% | 3.6% |
| LASSO, RF, GB, or NN | 56.2% | | | | | | | | | |
| OLS | 0% | -369% | -726% | -979% | -1,552% | -2,515% | -4,041% | -6,120% | -8,023% | |

This table reports the fraction of the S&P 500 stocks with positive $R^2_{oos}$ and the distribution of $R^2_{oos}$. $R^2_{oos}$ measures the fraction of daily return variations that can be explained by the out-of-sample return forecasts by foreign signals. Foreign signals are lagged weekly returns of leading stock market indices in 47 non-U.S. markets during the previous four weeks. Return forecasts are based on least absolute shrinkage and selection operator (LASSO) regression, random forest (RF), gradient boosted tree (GB), neural network (NN), and ordinary least squares regression (OLS). We also report the average $R^2_{oos}$ based on those stock-quarters with positive $R^2_{oos}$, and the fraction of the S&P 500 stocks with positive $R^2_{oos}$ based on return forecasts by at least one of the LASSO, RF, GB, or NN algorithms (LASSO, RF, GB, or NN).

## Table 4: Out-of-sample $R^2_{oos}$, Multinational and Domestic Firms, and Industries

| | Fraction of Stocks with $R^2_{oos}>0$ | Mean $R^2_{oos}$ (Stocks with $R^2_{oos}>0$) | | | |
|---|---|---|---|---|---|
| | | LASSO | RF | GB | NN |
| Multinational firms | 56.3% | 1.6% | 3.1% | 2.0% | 3.5% |
| Domestic firms | 56.0% | 1.6% | 3.0% | 2.0% | 3.6% |
| *Fama and French 12 Industry* | | | | | |
| Consumer Nondurables | 55.9% | 1.6% | 3.0% | 2.0% | 3.6% |
| Consumer Durables | 59.3% | 1.3% | 2.8% | 1.8% | 3.1% |
| Manufacturing | 55.0% | 1.5% | 3.3% | 2.0% | 3.4% |
| Oil, Gas, and Coal Extraction and Products | 56.6% | 1.6% | 2.7% | 1.9% | 3.5% |
| Chemicals | 58.6% | 1.6% | 2.9% | 2.0% | 3.4% |
| Computers, Software, and Electronic Equipment | 60.1% | 1.6% | 3.0% | 1.9% | 3.2% |
| Telephone and Television Transmission | 53.4% | 1.5% | 3.1% | 1.9% | 4.0% |
| Utilities | 48.8% | 1.6% | 3.1% | 2.1% | 2.8% |
| Wholesale and Retail | 56.8% | 1.6% | 3.0% | 2.0% | 3.6% |
| Healthcare | 54.8% | 1.4% | 2.9% | 2.0% | 4.3% |
| Finance | 54.4% | 1.6% | 3.0% | 2.1% | 3.8% |
| Other | 58.8% | 1.7% | 3.1% | 2.0% | 3.8% |

This table reports the fraction of the S&P 500 stocks with positive $R^2_{oos}$ and the average $R^2_{oos}$ for stocks with $R^2_{oos}>0$. We report the results for multinational firms, domestic firms, and firms in each of the Fama and French 12 industries. Multinational firms are those with more than 5% of total sales from non-U.S. markets. $R^2_{oos}$ measures the fraction of daily return variations that can be explained by the return forecasts by foreign signals. Foreign signals are lagged weekly returns of leading stock market indices in 47 non-U.S. markets during the previous four weeks. Return forecasts are based on least absolute shrinkage and selection operator (LASSO) regression, random forest (RF), gradient boosted tree (GB), neural network (NN), and ordinary least squares regression (OLS).

## Table 5: Signal Importance in Each Lagged Week

Panel A: Lagged week

|  | (1) LASSO | (2) RF | (3) GB | (4) NN |
|---|---|---|---|---|
| Lagged week | -0.0443*** | -0.0550*** | -0.0092*** | -0.0261*** |
|  | (-24.63) | (-49.54) | (-14.23) | (-5.99) |
| Constant | 0.2737*** | 1.0663*** | 0.3507*** | 3.6720*** |
|  | (47.92) | (319.13) | (191.38) | (258.52) |
| Observations | 873,448 | 805,204 | 1,333,296 | 286,700 |
| Adjusted $R^2$ | 0.0008 | 0.0027 | 0.0001 | 0.0001 |

Panel B: Exp (Lagged week)

|  | (1) LASSO | (2) RF | (3) GB | (4) NN |
|---|---|---|---|---|
| Exp (Lagged week) | -0.0021*** | -0.0027*** | -0.0004*** | -0.0007*** |
|  | (-24.84) | (-46.06) | (-11.58) | (-3.11) |
| Constant | 0.2065*** | 0.9865*** | 0.3364*** | 3.6224*** |
|  | (63.71) | (469.59) | (302.91) | (310.27) |
| Observations | 873,448 | 805,204 | 1,333,296 | 286,700 |
| Adjusted $R^2$ | 0.0005 | 0.0022 | 0.0001 | 0.0000 |

In this table, we regress signal importance on the lagged week indicator. For LASOO, signal importance is the absolute value of the coefficient on each signal. Signal importance is the SHAP value for RF, GB, and NN. We multiply the original signal importance values by $10^4$. Foreign signals are lagged weekly returns of leading stock market indices in 47 non-U.S. markets during the previous four weeks. In Panel A, Lagged week takes the value of $i$ for lagged $i$-week foreign signals. For example, it is 1 for lagged 1-week foreign signals and 4 for lagged 4-week foreign signals. The analyses are based on observations of the S&P 500 stocks with $R^2_{oos}>0$ by each algorithm. In Panel B, we use the natural exponential function of the lagged week indicator, exp (lagged week), as the independent variable. T-statistics are reported in parentheses. The standard errors are clustered by stock × quarter × country. *, **, and *** denote statistical significance at the 10%, 5%, and 1% level, respectively.

## Table 6: Factors Affecting Signal Importance in Each Lagged Week

|  | (1) LASSO | (2) RF | (3) GB | (4) NN |
|---|---|---|---|---|
| Exp (Lagged week) | -0.0009* | -0.0013*** | -0.0001 | 0.0027*** |
|  | (-1.86) | (-5.33) | (-0.39) | (3.40) |
| Exp (Lagged week) × news coverage of a market | 0.0058** | -0.0044** | -0.0078*** | -0.0271*** |
|  | (2.46) | (-1.96) | (-5.70) | (-3.77) |
| Exp (Lagged week) × developed market | 0.0022*** | -0.0005*** | -0.0006*** | -0.0050*** |
|  | (12.65) | (-4.00) | (-8.19) | (-10.05) |
| Exp (Lagged week) × foreign institutional ownership | -0.0104*** | -0.0104*** | -0.0033*** | 0.0139* |
|  | (-3.98) | (-5.33) | (-2.92) | (1.78) |
| Exp (Lagged week) × institutional ownership | 0.0009* | 0.0004 | 0.0007*** | -0.0003 |
|  | (1.96) | (1.12) | (3.41) | (-0.28) |
| Exp (Lagged week) × share turnover | -0.0011*** | -0.0002** | -0.0001** | -0.0004*** |
|  | (-4.98) | (-2.35) | (-2.05) | (-3.50) |
| Exp (Lagged week) × multinational firm | -0.0000 | -0.0005*** | 0.0000 | -0.0013** |
|  | (-0.16) | (-3.31) | (0.38) | (-2.20) |
| News coverage of a market | -0.7685*** | -0.7979*** | -0.2086*** | -2.7275*** |
|  | (-8.60) | (-10.31) | (-4.49) | (-10.30) |
| Developed market | -0.1670*** | -0.1723*** | -0.0744*** | 0.2813*** |
|  | (-24.65) | (-42.03) | (-31.90) | (18.15) |
| Foreign institutional ownership | -0.5553*** | -0.2127** | -0.1051* | 0.3415 |
|  | (-4.04) | (-2.05) | (-1.89) | (0.81) |
| Institutional ownership | 0.0287 | 0.0420** | 0.0015 | 0.1875** |
|  | (1.22) | (2.43) | (0.16) | (2.51) |
| Share turnover | 0.1631*** | 0.1173*** | 0.0446*** | 0.0117 |
|  | (10.58) | (25.86) | (18.74) | (1.60) |
| Multinational firm | 0.0097 | 0.0240*** | 0.0023 | 0.0792** |
|  | (0.77) | (3.19) | (0.54) | (1.97) |
|  |  |  |  |  |
| Quarter FE | Y | Y | Y | Y |
| Stock FE | Y | Y | Y | Y |
| Observations | 873,448 | 805,204 | 1,333,296 | 286,700 |
| Adjusted $R^2$ | 0.0221 | 0.0873 | 0.0236 | 0.4224 |

In this table, we regress signal importance on lagged week indicator. For LASOO, signal importance is the absolute value of the coefficient on each signal. Signal importance is the SHAP value for RF, GB, and NN. We multiply the original signal importance values by $10^{\wedge}4$. Foreign signals are lagged weekly returns of leading stock market indices in 47 non-U.S. markets during the previous four weeks. Lagged week takes the value of $i$ for lagged $i$-week foreign signals. For example, it is 1 for lagged 1-week foreign signals and 4 for lagged 4-week foreign signals. Exp (Lagged week) is the natural exponential function of Lagged week. News coverage of a market is the number of news stories regarding entities in one foreign market during the period to train one model scaled by the total number of news stories in the same time period. Data on news stories are from RavenPack. Developed market is a dummy variable, taking the value of 1 for signals from developed markets. Foreign institutional ownership is the percentage of a stock's shares outstanding held by foreign institutional investors. Institutional ownership is the percentage of a stock's shares outstanding held by institutional investors. Share turnover is trading volume divided by shares outstanding. Multinational firm is a dummy variable, taking the value of 1 for firms with more than 5% of total sales from foreign markets. The analyses are based on observations of the S&P 500 stocks with $R^2_{oos}>0$ by each algorithm. Fixed effects are included where indicated. T-statistics are reported in parentheses. The standard errors are clustered by stock × quarter × country. *, **, and *** denote statistical significance at the 10%, 5%, and 1% level, respectively.

# Table 7: Signal Importance in Each Lagged Week Over Time

Panel A: Early vs. Later Period

| | (1) LASSO | | (2) RF | | (3) GB | | (4) NN | |
|---|---|---|---|---|---|---|---|---|
| Exp (Lagged week) | -0.0012*** | -0.0021*** | -0.0016*** | -0.0010*** | -0.0001 | 0.0005*** | 0.0005** | 0.0018*** |
|  | (-13.95) | (-13.69) | (-21.53) | (-7.78) | (-1.21) | (6.93) | (2.33) | (5.89) |
| Exp (Lagged week) × later period | -0.0018*** | -0.0027*** | -0.0023*** | -0.0030*** | -0.0008*** | -0.0012*** | -0.0033*** | 0.0011 |
|  | (-10.50) | (-8.64) | (-19.10) | (-15.33) | (-11.12) | (-10.40) | (-5.71) | (1.04) |
| Exp (Lagged week) × later period × developed market |  | 0.0015*** |  | 0.0013*** |  | 0.0007*** |  | -0.0080*** |
|  |  | (4.27) |  | (5.25) |  | (5.11) |  | (-6.56) |
| Exp (Lagged week) × developed market |  | 0.0016*** |  | -0.0012*** |  | -0.0010*** |  | -0.0024*** |
|  |  | (8.77) |  | (-7.49) |  | (-11.41) |  | (-5.83) |
| Later period × developed market |  | -0.1128*** |  | -0.0387*** |  | -0.0158*** |  | 0.2344*** |
|  |  | (-8.03) |  | (-4.86) |  | (-3.46) |  | (6.27) |
| Developed market |  | -0.1248*** |  | -0.1638*** |  | -0.0700*** |  | 0.1628*** |
|  |  | (-19.45) |  | (-33.47) |  | (-25.87) |  | (13.63) |
| Quarter FE | Y | Y | Y | Y | Y | Y | Y | Y |
| Stock FE | Y | Y | Y | Y | Y | Y | Y | Y |
| Observations | 873,448 | 873,448 | 805,204 | 805,204 | 1,333,296 | 1,333,296 | 286,700 | 286,700 |
| Adjusted $R^2$ | 0.0166 | 0.0182 | 0.0759 | 0.0827 | 0.0190 | 0.0220 | 0.4215 | 0.4222 |

Panel B: Interaction with Year Variable

| | (1) LASSO | | (2) RF | | (3) GB | | (4) NN | |
|---|---|---|---|---|---|---|---|---|
| Exp (Lagged week) | -0.0023*** | -0.0037*** | -0.0030*** | -0.0029*** | -0.0005*** | -0.0002*** | -0.0004 | 0.0031*** |
| | (-23.20) | (-21.21) | (-47.41) | (-27.95) | (-12.95) | (-3.28) | (-1.21) | (5.17) |
| Exp (Lagged week) × year | -0.0003*** | -0.0004*** | -0.0005*** | -0.0007*** | -0.0001*** | -0.0002*** | 0.0003*** | 0.0008*** |
| | (-8.44) | (-7.86) | (-20.01) | (-18.81) | (-8.12) | (-9.29) | (3.15) | (3.94) |
| Exp (Lagged week) × year × developed market | | 0.0003*** | | 0.0004*** | | 0.0002*** | | -0.0008*** |
| | | (5.00) | | (9.18) | | (6.11) | | (-3.84) |
| Exp (Lagged week) × developed market | | 0.0026*** | | -0.0003* | | -0.0005*** | | -0.0062*** |
| | | (12.62) | | (-1.92) | | (-6.84) | | (-9.22) |
| Year × developed market | | -0.0210*** | | -0.0194*** | | -0.0057*** | | 0.0243*** |
| | | (-8.34) | | (-12.49) | | (-6.45) | | (3.87) |
| Developed market | | -0.1957*** | | -0.1964*** | | -0.0823*** | | 0.2753*** |
| | | (-23.90) | | (-44.92) | | (-32.41) | | (13.66) |
| | | | | | | | | |
| Quarter FE | Y | Y | Y | Y | Y | Y | Y | Y |
| Stock FE | Y | Y | Y | Y | Y | Y | Y | Y |
| Observations | 873,448 | 873,448 | 805,204 | 805,204 | 1,333,296 | 1,333,296 | 286,700 | 286,700 |
| Adjusted $R^2$ | 0.0166 | 0.0182 | 0.0760 | 0.0831 | 0.0190 | 0.0220 | 0.4215 | 0.4220 |

In this table, we regress signal importance on the lagged week indicator. For LASOO, signal importance is the absolute value of the coefficient on each signal. Signal importance is the SHAP value for RF, GB, and NN. We multiply the original signal importance values by $10^4$. Lagged week takes the value of $i$ for lagged $i$-week foreign signals. For example, it is 1 for lagged 1-week foreign signals and 4 for lagged 4-week foreign signals. Exp (Lagged week) is the natural exponential function of Lagged week. In Panel A, Later period is a dummy variable, taking the value of 1 for signals from models with training estimation windows ending in or after 2017. In Panel B, Year for each signal is the year of the last quarter in the corresponding model's estimation window minus 2017. Developed market is a dummy variable, taking the value of 1 for signals from developed markets. The analyses are based on observations of the S&P 500 stocks with $R^2_{oos}$ >0 by each algorithm. Fixed effects are included where indicated. T-statistics are reported in parentheses. *The coefficients on Later period and Year are absorbed by the Quarter fixed effects.* The standard errors are clustered by stock × quarter × country. *, **, and *** denote statistical significance at the 10%, 5%, and 1% level, respectively.

Table 8: COVID-19 and Signal Importance in Each Lagged Week

|  | (1)<br>LASSO | (2)<br>RF | (3)<br>GB | (4)<br>NN |
|---|---|---|---|---|
| Exp (Lagged week) | -0.0010*** | -0.0017*** | -0.0003*** | -0.0007*** |
|  | (-15.67) | (-28.39) | (-9.46) | (-2.96) |
| Exp (Lagged week) × COVID-19 | -0.0119*** | -0.0104*** | -0.0011*** | -0.0025* |
|  | (-17.96) | (-39.79) | (-5.31) | (-1.88) |
| COVID-19 | 0.7880*** | 0.6695*** | 0.2489*** | -0.5536*** |
|  | (26.43) | (64.65) | (36.01) | (-5.24) |
|  |  |  |  |  |
| Stock FE | Y | Y | Y | Y |
| Observations | 873,448 | 805,204 | 1,333,296 | 286,700 |
| Adjusted $R^2$ | 0.0164 | 0.0660 | 0.0157 | 0.1880 |

In this table, we study the impact of the COVID-19 outbreak on signal importance. COVID-19 is a dummy variable, taking the value of 1 for signals from models with a training period including 2020Q1. We regress signal importance on the lagged week indicator, COVID-19 dummy, and their interactions. Lagged week takes the value of $i$ for lagged $i$-week foreign signals. For example, it is 1 for lagged 1-week foreign signals and 4 for lagged 4-week foreign signals. Exp(Lagged week) is the natural exponential function of Lagged week. Foreign signals are lagged weekly returns of leading stock market indices in 47 non-U.S. markets during the previous four weeks. For LASOO, signal importance is the absolute value of the coefficient. Signal importance is the SHAP value for RF, GB, and NN. We multiply the original signal importance values by $10^{\wedge}4$. The analyses are based on observations of the S&P 500 stocks with $R^2_{oos}$>0 by each algorithm. Fixed effects are included where indicated. T-statistics are reported in parentheses. The standard errors are clustered by stock × quarter × country. *, **, and *** denote statistical significance at the 10%, 5%, and 1% level, respectively.

## Table 9: The Five Most Important Markets in Each Year

| Year | 2012 | 2013 | 2014 | 2015 | 2016 | 2017 | 2018 | 2019 | 2020 | 2021 |
|---|---|---|---|---|---|---|---|---|---|---|
| Rank | | | | | LASSO | | | | | |
| 1 | Egypt | Denmark | Israel | Switzerland | China | Egypt | Australia | Egypt | Kuwait | Japan |
| 2 | Mexico | Egypt | Mexico | Egypt | Kuwait | UK | Kuwait | Ireland | Mexico | Saudi Arabia |
| 3 | Qatar | Colombia | Greece | Greece | Australia | Denmark | Mexico | Mexico | Russia | Thailand |
| 4 | Thailand | Switzerland | Egypt | Singapore | Greece | Brazil | Egypt | China | Thailand | Peru |
| 5 | Turkey | India | UAE | New Zealand | Russia | Kuwait | Denmark | Singapore | Egypt | UAE |
| | | | | | RF | | | | | |
| 1 | Thailand | Denmark | Ireland | Singapore | Kuwait | Australia | Egypt | India | Israel | New Zealand |
| 2 | Mexico | Colombia | Indonesia | Malaysia | China | Denmark | India | Egypt | Mexico | UAE |
| 3 | Qatar | New Zealand | Greece | Hong Kong, China | Qatar | Saudi Arabia | Qatar | Thailand | Brazil | India |
| 4 | Indonesia | Greece | Thailand | Egypt | Saudi Arabia | South Africa | Denmark | Mexico | Russia | Mexico |
| 5 | Israel | Israel | UAE | Peru | UAE | Kuwait | Mexico | China | Australia | Kuwait |
| | | | | | GB | | | | | |
| 1 | Qatar | Denmark | Mexico | Singapore | China | Denmark | Australia | Egypt | Qatar | Denmark |
| 2 | Turkey | New Zealand | Hungary | Egypt | Peru | Australia | Israel | India | New Zealand | Norway |
| 3 | Thailand | Colombia | Kuwait | Hong Kong, China | Kuwait | Saudi Arabia | Egypt | Chile | Kuwait | Malaysia |
| 4 | Denmark | India | Indonesia | Qatar | New Zealand | Philippines | UAE | China | Russia | Kuwait |
| 5 | Mexico | Hungary | China | China | Russia | Colombia | Philippines | Singapore | Colombia | Peru |
| | | | | | NN | | | | | |
| 1 | Greece | Japan | Kuwait | Denmark | Norway | Germany | Germany | France | Denmark | Belgium |
| 2 | Saudi Arabia | Belgium | Qatar | Germany | Germany | Belgium | Chile | Germany | Israel | Colombia |
| 3 | Germany | Chile | Mexico | Malaysia | Finland | Spain | Indonesia | Chile | Turkey | Kuwait |
| 4 | Spain | Kuwait | Saudi Arabia | Turkey | Portugal | Denmark | Czech Republic | Italy | Poland | Indonesia |
| 5 | Qatar | Canada | Chile | Chile | Israel | France | Mexico | Poland | Kuwait | Chile |

This table reports the five foreign markets with the highest average signal importance in each year. We rank non-U.S. markets based on the average signal importance of foreign signals from a market in a year. Foreign signals are lagged weekly returns of leading stock market indices in 47 non-U.S. markets during the previous four weeks. For LASOO, the signal importance is the absolute value of the coefficient. The signal importance is the SHAP value for RF, GB, and NN.

## Table 10: Return Forecasts by a Pool of ≈13,000 Candidate Foreign Signals

| | Fraction of Stocks with $R^2_{oos}>0$ | $R^2_{oos}$ Distribution | | | | | | | | Mean (Stocks with $R^2_{oos}>0$) |
|---|---|---|---|---|---|---|---|---|---|---|
| | | 99th | 95th | 90th | 75th | 50th | 25th | 10th | 5th | |
| LASSO_13,000 | 11.9% | 3.9% | 1.4% | 0.3% | 0.0% | -0.2% | -1.8% | -5.4% | -11.9% | 1.6% |

This table analyzes the performance of return forecasts based on a pool of ≈ 13,000 candidate foreign signals. This pool includes lagged weekly returns of individual stocks and stock markets in 47 non-U.S. markets during the previous four weeks. Individual stock return signals include all the stocks with a top 10% market capitalization in each market. To predict the daily returns of the S&P 500 stocks, we use the least absolute shrinkage and selection operator (LASSO). We report the fraction of the S&P 500 stocks with positive $R^2_{oos}$ and the distribution of $R^2_{oos}$. $R^2_{oos}$ measures the fraction of daily return variations that can be explained by the out-of-sample return forecasts by foreign signals.

Internet Appendix I

Foreign Signal Radar

**Figure A1: An Example of the Relationships between Foreign Signals and Signal Importance**

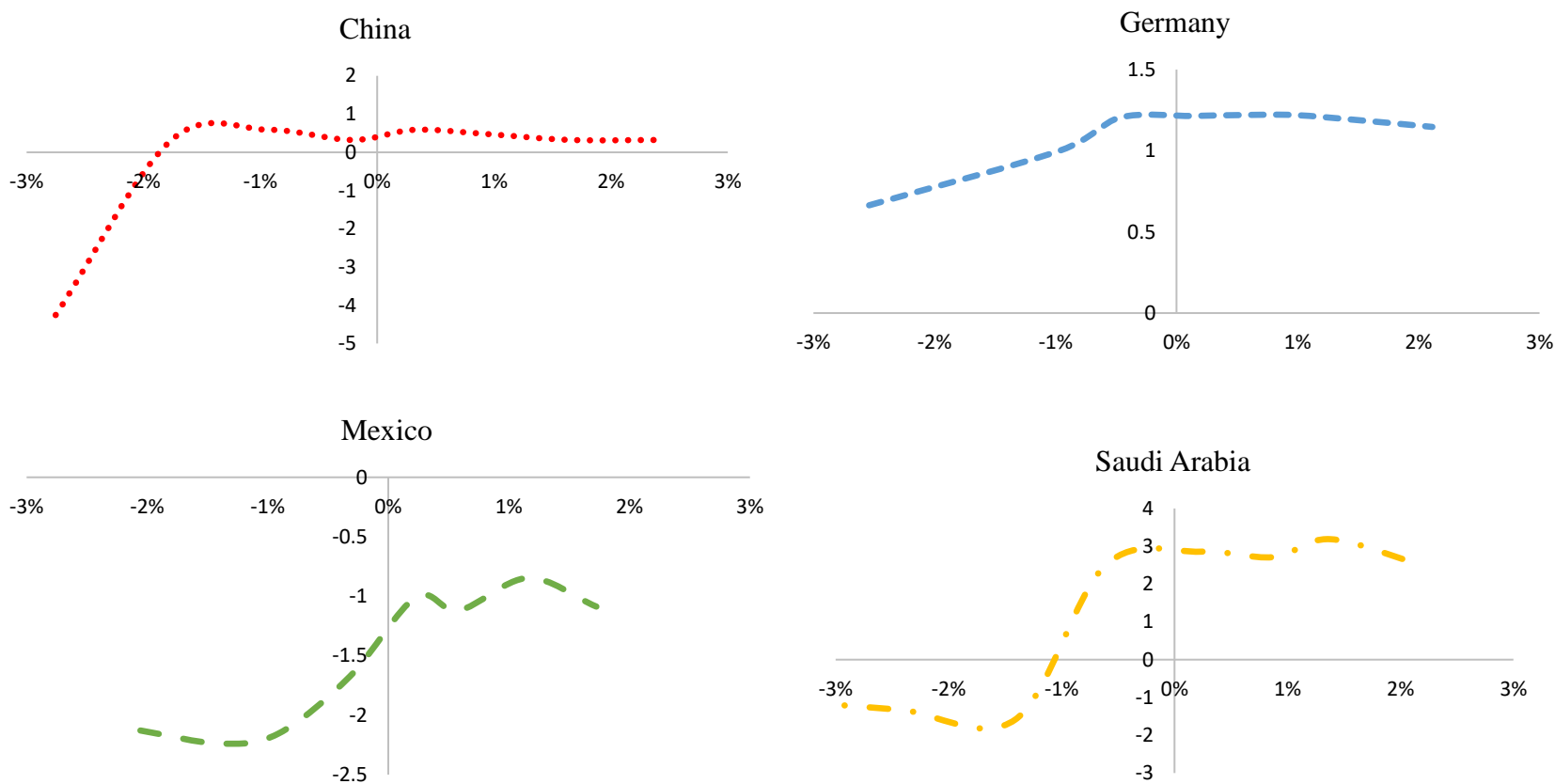

This figure shows the relationships between foreign signals and signal importance for Apple Inc. The horizontal axis shows the lagged 1-week market returns of China, Germany, Mexico, and Saudi Arabia. The vertical axis shows the signal importance. We measure signal importance using SHAP values. SHAP values reflect a signal's incremental contribution to the model's predictions. We multiply the original SHAP values by $10^{4}$. The algorithm we use is RF, and the model is trained on data from 2019Q2 to 2020Q1.

**Figure A2: The Cumulative Returns of Market Timing Strategies**

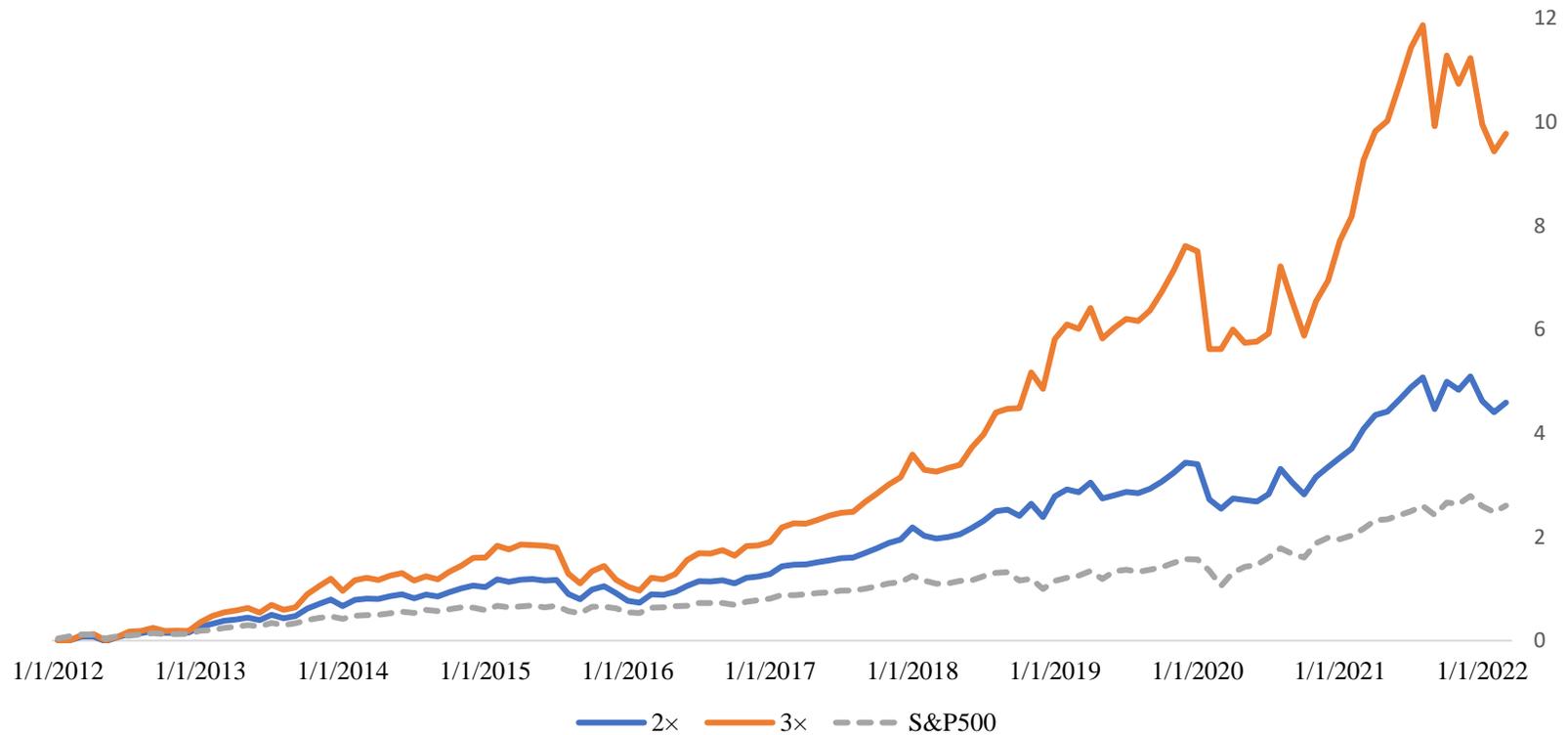

This figure shows the cumulative returns of market timing strategies using return forecasts by foreign signals. We compute bottom-up predicted returns of the S&P500 index by aggregating individual stock return predictions using foreign signals and four algorithms, LASSO, RF, GB, and NN. Each day, the strategy has 200% (300%) exposure to the S&P500 index if all four algorithms predict the S&P500 return to be positive. The strategy has -100% exposure to the S&P 500 index if all four algorithms predict the S&P500 return to be negative. In other cases, the strategy has 100% exposure to the S&P500 index. 2× (3×) indicates 200% (300%) exposure to the S&P500 index when all models' predictions are positive. We present the cumulative returns of the market timing strategies in excess of the daily one-month U.S. Treasury yield.

## Table A1: Sharpe Ratio, Maximum 1-quarter Loss, and Portfolio Turnover

|              | LASSO  | RF     | GB     | NN     | Comb   | OLS    | S&P 500 |
|--------------|--------|--------|--------|--------|--------|--------|---------|
|              | (1)    | (2)    | (3)    | (4)    | (5)    | (6)    | (7)     |
| Sharpe Ratio | 1.08   | 1.12   | 1.18   | 1.18   | 1.19   | 0.65   | 0.81    |
| Max 1Q loss  | -24.9% | -24.5% | -25.1% | -26.3% | -25.1% | -43.4% | -20.3%  |
| Turnover     | 32.8%  | 64.2%  | 55.8%  | 40.4%  | 42.2%  | 40.6%  |         |

This table reports the Sharpe ratio, maximum 1-quarter loss, and portfolio turnover for the portfolios of stocks with top 5% return forecasts. To predict the daily returns of the S&P 500 stocks, we use foreign signals and various machine learning algorithms, least absolute shrinkage and selection operator (LASSO) regression, random forest (RF), gradient boosted tree (GB), neural network (NN), and ordinary least squares regression (OLS). Foreign signals are lagged weekly returns of leading stock market indices in 47 non-U.S. markets during the previous four weeks. We report the annualized Sharpe ratio. Max 1Q loss is the most extreme negative return in a quarter. Turnover shows the average percentage of portfolio assets that are updated in a day. Column (7) reports the results for the S&P 500 index.

**Table A2: The Distribution of the S&P 500 Stocks across Multinational and Domestic Firms and Industries**

|  | Fraction of the S&P 500 Stocks in Each Category |
|---|---|
| Multinational firms | 21.6% |
| Domestic firms | 78.4% |
| **Fama and French 12 Industry** | |
| Consumer Nondurables | 6.6% |
| Consumer Durables | 2.0% |
| Manufacturing | 8.3% |
| Oil, Gas, and Coal Extraction and Products | 5.5% |
| Chemicals | 3.7% |
| Computers, Software, and Electronic Equipment | 14.6% |
| Telephone and Television Transmission | 3.1% |
| Utilities | 6.7% |
| Wholesale and Retail | 10.4% |
| Healthcare | 8.0% |
| Finance | 18.4% |
| Other | 12.8% |

This table presents the fraction of the S&P 500 stocks from multinational and domestic firms, and firms in each of the Fama and French 12 industries. Multinational firms are those with more than 5% of total sales from non-U.S. markets.

## Table A3: Signal Importance in Each Lagged Week, With Fixed Effects

Panel A: Lagged week

|  | (1) LASSO | (2) RF | (3) GB | (4) NN |
|---|---|---|---|---|
| Lagged week | -0.0443*** | -0.0550*** | -0.0092*** | -0.0261*** |
|  | (-24.62) | (-49.52) | (-14.23) | (-5.99) |
| Quarter FE | Y | Y | Y | Y |
| Stock FE | Y | Y | Y | Y |
| Observations | 873,448 | 805,204 | 1,333,296 | 286,700 |
| Adjusted $R^2$ | 0.0168 | 0.0761 | 0.0190 | 0.4215 |

Panel B: Exp (Lagged week)

|  | (1) LASSO | (2) RF | (3) GB | (4) NN |
|---|---|---|---|---|
| Exp (Lagged week) | -0.0021*** | -0.0027*** | -0.0004*** | -0.0007*** |
|  | (-24.83) | (-46.04) | (-11.58) | (-3.10) |
| Quarter FE | Y | Y | Y | Y |
| Stock FE | Y | Y | Y | Y |
| Observations | 873,448 | 805,204 | 1,333,296 | 286,700 |
| Adjusted $R^2$ | 0.0165 | 0.0755 | 0.0190 | 0.4214 |

In this table, we regress signal importance on the lagged week indicator. For LASOO, signal importance is the absolute value of the coefficient. Signal importance is the SHAP value for RF, GB, and NN. We multiply the original signal importance values by $10^{4}$. Foreign signals are lagged weekly returns of leading stock market indices in 47 non-U.S. markets during the previous four weeks. In Panel A, Lagged week takes the value of $i$ for lagged $i$-week foreign signals. For example, it is 1 for lagged 1-week foreign signals and 4 for lagged 4-week foreign signals. Exp (Lagged week) is the natural exponential function of Lagged week. The analyses are based on observations of the S&P 500 stocks with $R^2_{oos}>0$ by each algorithm. In Panel B, we use the natural exponential function of lagged week, exp(lagged week), as the independent variable. T-statistics are reported in parentheses. The standard errors are clustered by stock × quarter × country. *, **, and *** denote statistical significance at the 10%, 5%, and 1% level, respectively.

**Table A4: Country Characteristics and Signal Importance**

|  | (1) LASSO | (2) RF | (3) GB | (4) NN. |
|---|---|---|---|---|
| International trade | 0.1540*** | 0.1746*** | 0.0466*** | -0.4097*** |
|  | (7.97) | (15.79) | (7.98) | (-11.21) |
| Economic uncertainty | -0.0310*** | -0.1262*** | -0.0184*** | -0.0671*** |
|  | (-3.23) | (-19.73) | (-5.25) | (-2.90) |
| Geopolitical risk | -0.1514*** | -0.2768*** | -0.0768*** | -1.0637*** |
|  | (-4.37) | (-11.40) | (-5.74) | (-10.98) |
| Foreign sales | 0.0110 | 0.0494** | 0.0008 | -0.0248 |
|  | (0.41) | (2.25) | (0.07) | (-0.37) |
|  |  |  |  |  |
| Quarter FE | Y | Y | Y | Y |
| Stock FE | Y | Y | Y | Y |
| Lagged week FE | Y | Y | Y | Y |
| Observations | 706,192 | 651,016 | 1,077,984 | 231,800 |
| Adjusted $R^2$ | 0.0148 | 0.0762 | 0.0172 | 0.4448 |

In this table, we regress signal importance of foreign signals on various country characteristics. For LASOO, signal importance is the absolute value of the coefficient. Signal importance is the SHAP value for RF, GB, and NN. Foreign signals are lagged weekly returns of leading stock market indices in 47 non-U.S. markets during the previous four weeks. We multiply the original signal importance values by $10^{\wedge}4$. International trade is the U.S. trade in goods with a foreign country in $trillion. Geopolitical risk is based on the geopolitical risk index constructed by Caldara and Iacoviello (2022). Economic uncertainty is based on the world uncertainty index constructed by Ahir, Bloom, and Furceri (2022). Foreign sales is a dummy variable, taking the value of 1 if a firm has at least 5% of its total sales in a foreign market. The analyses are based on observations of the S&P 500 stocks with $R^2_{oos}>0$ by each algorithm. Fixed effects are included where indicated. T-statistics are reported in parentheses. The standard errors are clustered by stock × quarter × country. *, **, and *** denote statistical significance at the 10%, 5%, and 1% level, respectively.

## Table A5: Stock-Specific and Time-Varying Relationships

Panel A: Stock-specific Relationships

|  | (1)<br>LASSO | (2)<br>RF | (3)<br>GB | (4)<br>NN |
|---|---|---|---|---|
| Country × lagged week × quarter FE | Y | Y | Y | Y |
| Observations | 873,448 | 805,204 | 1,333,296 | 286,700 |
| $R^2$ | 0.1176 | 0.1425 | 0.1140 | 0.7431 |

Panel B: Time-varying Relationships

|  | (1)<br>LASSO | (2)<br>RF | (3)<br>GB | (4)<br>NN |
|---|---|---|---|---|
| Stock × country × lagged week FE | Y | Y | Y | Y |
| Observations | 873,448 | 805,204 | 1,333,296 | 286,700 |
| $R^2$ | 0.1830 | 0.2246 | 0.1506 | 0.3069 |

In this table, we regress signal importance on fixed effects. For LASOO, signal importance is the absolute value of the coefficient. Signal importance is the SHAP value for RF, GB, and NN. We multiply the original signal importance values by $10^{\wedge}4$. Foreign signals are lagged weekly returns of leading stock market indices in 47 non-U.S. markets during the previous four weeks. Lagged week takes the value of $i$ for lagged $i$-week foreign signals. For example, it is 1 for lagged 1-week foreign signals and 4 for lagged 4-week foreign signals. The analyses are based on observations of the S&P 500 stocks with $R^2_{oos}>0$ by each algorithm. In Panel A, we regress signal importance on country × lagged week × quarter fixed effects. In Panel B, we regress signal importance on stock × country × lagged week fixed effects.

## Table A6: Predicting the S&P 500 Index Returns

Panel A: Market Timing Strategy

|  | (1) | (2) | (3) | (4) | (5) |
|---|---|---|---|---|---|
|  | 2× | 3× | S&P 500 | (1) - (3) | (2) - (3) |
| Mean return (bps) | 7.66*** | 10.94*** | 5.30*** | 2.35** | 5.64** |
|  | (2.78) | (3.00) | (2.59) | (1.98) | (2.56) |
| Sharpe ratio | 0.87 | 0.94 | 0.81 |  |  |
| Max 1Q loss | -20.1% | -25.7% | -20.3% |  |  |
| Turnover |  | 28.5% |  |  |  |

Panel B: Out-of-sample $R^2_{oos}$

| | (1) Fraction of Quarters with $R^2_{oos}>0$ | (2) | (3) | (4) | (5) | (6) |
|---|---|---|---|---|---|---|
| | | \multicolumn{4}{c}{$R^2_{oos}$ Statistics} | Mean $R^2_{oos}$ |
| | | Mean | Median | Max | Min | (Quarters with $R^2_{oos}>0$) |
| LASSO | 56% | 0.35% | 0.36% | 7.15% | -6.36% | 2.09% |
| RF | 39% | -1.28% | -1.03% | 7.47% | -13.51% | 3.21% |
| GB | 49% | 0.19% | -0.05% | 6.60% | -7.55% | 2.54% |
| NN | 0% | -302.35% | -61.96% | -5.47% | -6789.69% |  |
| LASSO, RF, GB, or NN | 68% |  |  |  |  |  |

In this table, we report the performance of market-timing strategies. We compute bottom-up forecasted daily returns of the S&P 500 index by aggregating individual stock return predictions using foreign signals and four algorithms, LASSO, RF, GB, and NN. Foreign signals are lagged weekly returns of leading stock market indices in 47 non-U.S. markets during the previous four weeks. In Panel A, each day, the strategy has 200% (300%) exposure to the S&P500 index if all four algorithms predict the S&P 500 return to be positive. The strategy has -100% exposure to the S&P 500 index if all four algorithms predict the S&P 500 return to be negative. In other cases, the strategy has 100% exposure to the S&P 500 index. 2× (3×) indicates 200% (300%) exposure to the S&P500 index when all models' predictions are positive. We report the average daily returns of the market timing strategies in excess of the daily one-month U.S. Treasury yield. We also report the annualized Sharpe ratio. Max 1Q loss is the most extreme negative return in a quarter. Turnover shows the average turnover in a day. The turnover is 100% in a day if the strategy changes the exposure to the S&P 500 index from the previous day. It is 0% if the exposure to the S&P 500 index changes from the previous day. In Panel B, we report the fraction of quarters with positive $R^2_{oos}$ and the distribution of $R^2_{oos}$. $R^2_{oos}$ measures the fraction of daily return variations of the S&P 500 index that can be explained by the return forecasts. We also report the fraction of quarters with positive $R^2_{oos}$ based on return forecasts by at least one of the LASSO, RF, GB, or NN algorithms (LASSO, RF, GB, or NN).

## Table A7: Signal Importance for a Pool of ≈13,000 Foreign Signals

Panel A: Fraction of Stocks with $R^2_{oos}>0$

|  | Fraction of Stocks with $R^2_{oos}>0$ |
|---|---|
| LASSO_13,000 | 11.9% |
| LASSO_188 or LASSO_13,000 | 29.0% |
| LASSO_188, RF_188, GB_188, NN_188, or LASSO_13,000 | 59.0% |

Panel B: Daily Portfolio Returns

|  | LASSO | | |
|---|---|---|---|
|  | (1) Top | (2) Bottom | (3) T-B |
| Mean return | 9.52*** | 3.93 | 5.59*** |
|  | (3.07) | (1.16) | (2.64) |
| Alpha | 2.68* | -2.44* | 5.12*** |
|  | (1.93) | (-1.77) | (2.79) |
| MKT | 1.19*** | 1.16*** | 0.03 |
|  | (47.84) | (33.06) | (0.93) |
| SMB | 0.41*** | 0.34*** | 0.07 |
|  | (11.03) | (8.12) | (1.32) |
| HML | 0.37*** | 0.20*** | 0.17*** |
|  | (9.60) | (4.83) | (3.43) |
| MOM | -0.04 | -0.67*** | 0.63*** |
|  | (-1.28) | (-19.73) | (10.94) |
| RMW | -0.09 | -0.21*** | 0.12 |
|  | (-1.44) | (-4.56) | (1.58) |
| CMA | -0.08 | 0.16** | -0.24** |
|  | (-1.02) | (2.14) | (-2.27) |
| Observations | 2,579 | 2,579 | 2,579 |
| Adjusted R² | 0.81 | 0.82 | 0.26 |

Panel C: Signals from Foreign Firms in the Same Industry as U.S. Firms

|  | (1) All | (2) Consumer Nondurables | (3) Consumer Durables | (4) Manufacturing | (5) Oil, Gas, and Coal Extraction and Products |
|---|---|---|---|---|---|
| Same industry | -0.0001 | 0.0035** | -0.0001 | -0.0005 | -0.0005 |
|  | (-0.46) | (2.21) | (-0.17) | (-1.23) | (-0.92) |
|  |  |  |  |  |  |
| Quarter FE | Y | Y | Y | Y | Y |
| Stock FE | Y | Y | Y | Y | Y |
| Lagged week FE | Y | Y | Y | Y | Y |
| Observations | 28,423,509 | 1,232,068 | 681,754 | 2,184,738 | 2,505,942 |
| Adjusted $R^2$ | 0.0003 | 0.0008 | 0.0001 | 0.0004 | 0.0003 |

|  | (5) Chemicals | (6) Computers, Software, and Electronic Equipment | (7) Telephone and Television Transmission | (8) Utilities | (9) Wholesale and Retail | (10) Healthcare | (11) Finance | (12) Other |
|---|---|---|---|---|---|---|---|---|
| Same industry | -0.0022*** | -0.0004** | 0.0047** | 0.0036* | 0.0006 | 0.0002 | 0.0002 | -0.0005** |
|  | (-6.85) | (-2.00) | (2.43) | (1.80) | (1.38) | (0.41) | (0.63) | (-2.09) |
|  |  |  |  |  |  |  |  |  |
| Quarter FE | Y | Y | Y | Y | Y | Y | Y | Y |
| Stock FE | Y | Y | Y | Y | Y | Y | Y | Y |
| Lagged week FE | Y | Y | Y | Y | Y | Y | Y | Y |
| Observations | 1,077,699 | 5,695,293 | 1,085,723 | 780,954 | 2,791,304 | 2,339,122 | 3,522,330 | 4,526,582 |
| Adjusted $R^2$ | 0.0004 | 0.0003 | 0.0003 | 0.0003 | 0.0005 | 0.0004 | 0.0006 | 0.0004 |

Panel D: Signal Importance in each Lagged Week

| | (1) | (2) |
|---|---|---|
| Exp (Lagged week) | -0.0032*** | -0.0017 |
| | (-18.39) | (-1.45) |
| Exp (Lagged week) × news coverage of a market | | -0.0587*** |
| | | (-10.04) |
| Exp (Lagged week) × developed market | | 0.0028*** |
| | | (7.62) |
| Exp (Lagged week) × foreign institutional ownership | | -0.0107** |
| | | (-2.46) |
| Exp (Lagged week) × institutional ownership | | 0.0008 |
| | | (0.87) |
| Exp (Lagged week) × share turnover | | -0.0002 |
| | | (-0.63) |
| Exp (Lagged week) × multinational firm | | -0.0002 |
| | | (-0.63) |
| Exp (Lagged week) × individual foreign stock signal | | -0.0009 |
| | | (-1.20) |
| Constant | 0.3787*** | -0.3392*** |
| | (57.49) | (-5.31) |
| Controls | | Y |
| Quarter FE | | Y |
| Stock FE | | Y |
| Observations | 30,918,310 | 30,918,310 |
| Adjusted R² | 0.0000 | 0.0004 |

Panel E: Fraction of Signals selected by LASSO

| | Fraction of signals with non-zero coefficients |
|---|---|
| LASSO_13,000 | 0.0538% |
| LASSO_188 | 2.24% |

This table analyzes the performance of return forecasts based on a pool of ≈13,000 candidate foreign signals. This pool includes lagged weekly returns of stock market indices and individual stocks with top 10% market cap in each of 47 non-U.S. markets during the previous four weeks. Panel A reports the out-of-sample $R^2_{oos}$. LASSO_13,000 indicates the statistics based on return forecasts by LASSO and the pool of ≈13,000 candidate foreign signals. LASSO_188, RF_188, GB_188, NN_188 indicate the statistics based on return forecasts by LASSO, RF, GB, or NN using the 188 foreign stock market return signals. Panel B presents the daily performance of portfolios formed on foreign signal return forecasts. Top refers to the portfolios of stocks with top 5% return forecasts. Bottom indicates the portfolios of stocks with bottom 5% return forecasts. We equally weight stocks in each portfolio. T-B is the portfolio that buys stocks with top 5% return forecasts and sells short stocks with bottom 5% return forecasts. Portfolios are rebalanced every day. Robust *t*-statistics are reported in parentheses. In Panel C, we regress signal importance on the same industry dummy variable. For LASOO, signal importance is the absolute value of the coefficient. We multiply the original signal importance values by 10^4. Same industry is a dummy, taking the value of 1 if the foreign firm and the U.S. firm are in the same industry. Industry is based on the Fama and French 12 industry classification. In Panel D, we regress signal importance on the lagged week indicator. Column (2) follows the settings in Table 6. Individual foreign stock signal is a dummy, taking the value of 1 for individual foreign stock return signals and 0 for foreign stock market return signals. *For brevity, we omit the coefficients on those characteristics of markets and stocks.* The analyses in Panels B and C are based on observations of the S&P 500 stocks with $R^2_{oos}>0$. Fixed effects are included where indicated. Fixed effects are included where indicated. T-statistics are reported in parentheses. The standard errors are clustered by stock × quarter × country. *, **, and *** denote statistical significance at the 10%, 5%, and 1% level, respectively. In Panel E, we report the average ratio of the number of signals with non-zero coefficients to the total number of candidate foreign signals. LASSO_13,000 indicates the statistics based on return forecasts by LASSO and the pool of ≈13,000 candidate foreign signals. LASSO_188 indicates the statistics based on return forecasts by LASSO using the 188 foreign stock market return signals.

## Table A8: Daily Performance of Portfolios Formed on Return Forecasts by Foreign Signals, Subperiod Analysis

Panel A: Year<=2016

|  | LASSO | | | RF | | | GB | | |
|---|---|---|---|---|---|---|---|---|---|
|  | (1) | (2) | (3) | (4) | (5) | (6) | (7) | (8) | (9) |
|  | Top | Bottom | T-B | Top | Bottom | T-B | Top | Bottom | T-B |
| Mean return | 10.07*** | 5.01 | 5.06* | 10.08*** | 4.78 | 5.31** | 10.10*** | 3.67 | 6.43** |
|  | (2.98) | (1.23) | (1.73) | (3.23) | (1.37) | (2.45) | (3.37) | (0.96) | (2.29) |
|  |  |  |  |  |  |  |  |  |  |
| Alpha | 3.14* | -1.73 | 4.87** | 3.29*** | -1.66 | 4.96** | 3.33*** | -2.81* | 6.14*** |
|  | (1.96) | (-0.99) | (2.08) | (2.61) | (-1.09) | (2.58) | (2.71) | (-1.75) | (3.17) |
| MKT | 1.21*** | 1.18*** | 0.02 | 1.17*** | 1.13*** | 0.04 | 1.14*** | 1.14*** | -0.00 |
|  | (50.05) | (42.54) | (0.71) | (59.68) | (49.77) | (1.24) | (59.79) | (46.79) | (-0.02) |
| SMB | 0.05 | 0.16*** | -0.12* | 0.07* | 0.14*** | -0.08 | 0.04 | 0.13*** | -0.08* |
|  | (1.11) | (3.23) | (-1.77) | (1.88) | (2.95) | (-1.33) | (1.25) | (2.87) | (-1.67) |
| HML | -0.17*** | 0.00 | -0.17** | -0.04 | 0.01 | -0.05 | -0.07 | 0.05 | -0.12* |
|  | (-2.71) | (0.07) | (-2.02) | (-0.97) | (0.18) | (-0.76) | (-1.52) | (0.88) | (-1.79) |
| MOM | -0.13*** | -0.94*** | 0.81*** | -0.13*** | -0.61*** | 0.49*** | 0.10*** | -0.88*** | 0.97*** |
|  | (-3.94) | (-24.98) | (16.42) | (-4.93) | (-17.24) | (11.57) | (4.23) | (-23.47) | (23.50) |
| RMW | -0.31*** | -0.29*** | -0.02 | -0.21*** | -0.12 | -0.09 | -0.23*** | -0.15* | -0.08 |
|  | (-4.18) | (-3.91) | (-0.23) | (-3.61) | (-1.16) | (-0.83) | (-4.37) | (-1.76) | (-0.86) |
| CMA | 0.20 | 0.29*** | -0.10 | 0.09 | 0.12 | -0.03 | 0.16* | 0.15* | 0.00 |
|  | (1.57) | (2.90) | (-0.62) | (1.05) | (1.27) | (-0.27) | (1.80) | (1.67) | (0.02) |
|  |  |  |  |  |  |  |  |  |  |
| Observations | 1,258 | 1,258 | 1,258 | 1,258 | 1,258 | 1,258 | 1,258 | 1,258 | 1,258 |
| Adjusted $R^2$ | 0.77 | 0.82 | 0.35 | 0.84 | 0.81 | 0.21 | 0.83 | 0.83 | 0.51 |

|  | NN | | | Comb | | | OLS | | |
|---|---|---|---|---|---|---|---|---|---|
|  | (10) | (11) | (12) | (13) | (14) | (15) | (16) | (17) | (18) |
|  | Top | Bottom | T-B | Top | Bottom | T-B | Top | Bottom | T-B |
| Mean return | 9.01*** | 3.13 | 5.89* | 9.82*** | 4.15 | 5.67** | 5.95* | 6.86** | -0.91 |
|  | (3.18) | (0.79) | (1.91) | (3.35) | (1.12) | (2.48) | (1.86) | (2.11) | (-0.39) |
| Alpha | 2.67** | -3.17* | 5.84*** | 3.11*** | -2.35* | 5.45*** | -0.39 | 0.58 | -0.98 |
|  | (2.07) | (-1.77) | (2.68) | (3.10) | (-1.75) | (3.66) | (-0.26) | (0.38) | (-0.42) |
| MKT | 1.10*** | 1.15*** | -0.05 | 1.15*** | 1.15*** | 0.00 | 1.11*** | 1.08*** | 0.03 |
|  | (52.76) | (39.29) | (-1.46) | (74.19) | (54.47) | (0.05) | (53.45) | (42.41) | (0.79) |
| SMB | 0.03 | 0.06 | -0.03 | 0.05* | 0.12*** | -0.08* | 0.10** | 0.05 | 0.05 |
|  | (0.88) | (1.19) | (-0.47) | (1.65) | (3.02) | (-1.70) | (2.41) | (1.35) | (0.74) |
| HML | -0.14*** | -0.12 | -0.02 | -0.10*** | -0.01 | -0.09* | -0.12** | -0.07 | -0.05 |
|  | (-3.01) | (-1.55) | (-0.22) | (-2.95) | (-0.25) | (-1.71) | (-2.18) | (-1.06) | (-0.58) |
| MOM | 0.14*** | -0.92*** | 1.06*** | -0.01 | -0.84*** | 0.83*** | -0.31*** | -0.42*** | 0.11** |
|  | (4.02) | (-17.25) | (16.92) | (-0.25) | (-24.80) | (24.34) | (-8.73) | (-12.92) | (2.04) |
| RMW | -0.02 | -0.27*** | 0.25** | -0.20*** | -0.21*** | 0.01 | -0.20*** | -0.32*** | 0.12 |
|  | (-0.41) | (-2.87) | (2.18) | (-4.31) | (-2.77) | (0.16) | (-2.88) | (-3.97) | (1.17) |
| CMA | 0.16** | 0.19 | -0.04 | 0.15** | 0.19** | -0.04 | 0.21** | 0.29** | -0.08 |
|  | (2.17) | (1.54) | (-0.24) | (2.15) | (2.40) | (-0.47) | (2.39) | (2.43) | (-0.49) |
| Observations | 1,258 | 1,258 | 1,258 | 1,258 | 1,258 | 1,258 | 1,258 | 1,258 | 1,258 |
| Adjusted $R^2$ | 0.80 | 0.79 | 0.50 | 0.88 | 0.87 | 0.57 | 0.77 | 0.77 | 0.01 |

Panel B: Year>=2017

|  | LASSO | | | RF | | | GB | | |
|---|---|---|---|---|---|---|---|---|---|
|  | (1) Top | (2) Bottom | (3) T-B | (4) Top | (5) Bottom | (6) T-B | (7) Top | (8) Bottom | (9) T-B |
| Mean return | 10.71** | 4.66 | 6.05* | 9.26** | 6.93 | 2.33 | 9.36** | 7.95 | 1.41 |
|  | (2.17) | (0.86) | (1.73) | (2.12) | (1.50) | (1.00) | (2.27) | (1.50) | (0.39) |
| Alpha | 4.78** | -1.68 | 6.46** | 3.34** | 0.82 | 2.51 | 3.45** | 2.00 | 1.45 |
|  | (2.19) | (-0.77) | (2.01) | (2.05) | (0.50) | (1.11) | (2.17) | (1.03) | (0.57) |
| MKT | 1.14*** | 1.21*** | -0.07 | 1.09*** | 1.11*** | -0.02 | 1.06*** | 1.14*** | -0.08 |
|  | (41.43) | (20.82) | (-1.00) | (49.98) | (30.47) | (-0.50) | (67.09) | (20.85) | (-1.41) |
| SMB | 0.44*** | 0.37*** | 0.07 | 0.26*** | 0.31*** | -0.05 | 0.20*** | 0.40*** | -0.21*** |
|  | (8.83) | (6.76) | (0.98) | (8.17) | (8.63) | (-1.03) | (5.90) | (7.56) | (-3.11) |
| HML | 0.37*** | 0.24*** | 0.13* | 0.25*** | 0.24*** | 0.01 | 0.29*** | 0.28*** | 0.01 |
|  | (6.90) | (5.15) | (1.72) | (6.75) | (6.74) | (0.25) | (8.77) | (6.27) | (0.15) |
| MOM | -0.06* | -0.51*** | 0.45*** | -0.14*** | -0.28*** | 0.15*** | 0.16*** | -0.53*** | 0.69*** |
|  | (-1.87) | (-10.68) | (6.61) | (-2.92) | (-10.11) | (3.68) | (7.27) | (-17.70) | (18.89) |
| RMW | -0.03 | -0.06 | 0.03 | -0.05 | 0.04 | -0.09 | -0.09** | 0.04 | -0.13** |
|  | (-0.43) | (-0.98) | (0.31) | (-1.11) | (1.00) | (-1.56) | (-2.24) | (0.73) | (-1.98) |
| CMA | 0.01 | 0.12 | -0.11 | 0.00 | 0.17*** | -0.16** | 0.08 | 0.08 | 0.00 |
|  | (0.09) | (1.53) | (-0.89) | (0.03) | (2.82) | (-2.00) | (1.19) | (0.94) | (0.00) |
| Observations | 1,321 | 1,321 | 1,321 | 1,321 | 1,321 | 1,321 | 1,321 | 1,321 | 1,321 |
| Adjusted $R^2$ | 0.80 | 0.83 | 0.13 | 0.86 | 0.87 | 0.06 | 0.85 | 0.85 | 0.46 |

|  | NN | | | Comb | | | OLS | | |
|---|---|---|---|---|---|---|---|---|---|
|  | (10) | (11) | (12) | (13) | (14) | (15) | (16) | (17) | (18) |
|  | Top | Bottom | T-B | Top | Bottom | T-B | Top | Bottom | T-B |
| Mean return | 10.36** | 4.21 | 6.14* | 9.92** | 5.94 | 3.98* | 6.19 | 7.76* | -1.57 |
|  | (2.45) | (0.88) | (1.90) | (2.36) | (1.22) | (1.67) | (1.28) | (1.71) | (-0.44) |
| Alpha | 4.22** | -1.53 | 5.75** | 3.95*** | -0.10 | 4.04** | -0.04 | 2.34 | -2.38 |
|  | (2.23) | (-0.76) | (2.03) | (2.97) | (-0.06) | (2.16) | (-0.02) | (1.12) | (-0.67) |
| MKT | 1.07*** | 1.07*** | 0.00 | 1.09*** | 1.13*** | -0.04 | 1.14*** | 1.02*** | 0.12** |
|  | (31.77) | (40.52) | (0.08) | (61.90) | (29.41) | (-1.12) | (24.25) | (31.12) | (2.44) |
| SMB | 0.14*** | 0.31*** | -0.17** | 0.26*** | 0.35*** | -0.09** | 0.33*** | 0.27*** | 0.06 |
|  | (2.64) | (7.18) | (-2.37) | (8.83) | (9.08) | (-2.00) | (5.43) | (6.09) | (0.74) |
| HML | 0.18*** | 0.21*** | -0.03 | 0.27*** | 0.24*** | 0.03 | 0.29*** | 0.23*** | 0.06 |
|  | (4.27) | (5.55) | (-0.58) | (8.65) | (7.40) | (0.69) | (5.65) | (5.91) | (0.77) |
| MOM | -0.02 | -0.44*** | 0.42*** | -0.01 | -0.44*** | 0.43*** | -0.07 | -0.32*** | 0.25 |
|  | (-0.29) | (-6.88) | (9.83) | (-0.50) | (-11.71) | (14.95) | (-0.98) | (-3.46) | (1.54) |
| RMW | -0.08 | 0.04 | -0.12 | -0.06* | 0.02 | -0.08 | -0.00 | -0.05 | 0.05 |
|  | (-1.48) | (0.74) | (-1.63) | (-1.66) | (0.37) | (-1.55) | (-0.06) | (-0.79) | (0.44) |
| CMA | 0.07 | -0.04 | 0.11 | 0.04 | 0.08 | -0.04 | -0.07 | 0.06 | -0.14 |
|  | (1.17) | (-0.54) | (1.16) | (0.72) | (1.33) | (-0.55) | (-0.78) | (0.82) | (-1.04) |
| Observations | 1,321 | 1,321 | 1,321 | 1,321 | 1,321 | 1,321 | 1,321 | 1,321 | 1,321 |
| Adjusted $R^2$ | 0.80 | 0.83 | 0.24 | 0.90 | 0.90 | 0.37 | 0.78 | 0.80 | 0.05 |

This table presents the daily performance of portfolios formed on return forecasts by foreign signals. To predict the daily returns of the S&P 500 stocks, we use foreign signals and various machine learning algorithms, least absolute shrinkage and selection operator (LASSO) regression, random forest (RF), gradient boosted tree (GB), neural network (NN), and ordinary least squares regression (OLS). Foreign signals are lagged weekly returns of leading stock market indices in 47 non-U.S. markets during the previous four weeks. Top refers to the portfolios of stocks with top 5% return forecasts. Bottom means the portfolios of stocks with bottom 5% return forecasts. We equally weight stocks in each portfolio. T-B is the portfolio that buys stocks with top 5% return forecasts and sells short stocks with bottom 5% return forecasts. Portfolios are rebalanced every day. We report the mean return and intercept (Alpha) and loadings from regressions of Top or Bottom portfolio returns in excess of the daily one-month U.S. Treasury yield on the Fama and French (1993, 2015) MKT, SMB, HML, RMW, CMA, and Carhart (1997) MOM factors. We also report the mean return and intercept (Alpha) and loadings from the regression of T-B portfolio returns on the same set of factors. In columns (13) to (15), we equally weight the returns of portfolios based on LASSO, RF, GB, and NN for Top, Bottom, or T-B, respectively. All results are in basis points. Panel A is based on observations before 2017. Panel B includes the sample from 2017 to 2022Q1. Robust *t*-statistics are reported in parentheses. *, **, and *** denote statistical significance at the 10%, 5%, and 1% level, respectively.

**Table A9: Daily Performance of Portfolios Formed on Return Forecasts by Foreign Signals, Global ex-U.S. Factors**

|  | LASSO | | | RF | | | GB | | |
|---|---|---|---|---|---|---|---|---|---|
|  | (1) | (2) | (3) | (4) | (5) | (6) | (7) | (8) | (9) |
|  | Top | Bottom | T-B | Top | Bottom | T-B | Top | Bottom | T-B |
| Alpha | 9.27*** | 4.12 | 5.15** | 8.74*** | 4.93** | 3.81** | 8.14*** | 5.20** | 2.95 |
|  | (3.60) | (1.52) | (2.32) | (3.94) | (2.12) | (2.40) | (3.74) | (2.02) | (1.43) |
| MKT | 0.81*** | 0.95*** | -0.14*** | 0.75*** | 0.83*** | -0.09*** | 0.74*** | 0.90*** | -0.16*** |
|  | (12.52) | (13.57) | (-2.90) | (13.69) | (14.07) | (-3.00) | (13.00) | (13.36) | (-4.17) |
| SMB | -0.61*** | -0.56*** | -0.05 | -0.70*** | -0.61*** | -0.10* | -0.65*** | -0.55*** | -0.10 |
|  | (-5.37) | (-5.02) | (-0.64) | (-7.54) | (-5.29) | (-1.65) | (-6.53) | (-4.68) | (-1.32) |
| HML | 0.07 | 0.34 | -0.27 | -0.00 | 0.20 | -0.20* | -0.07 | 0.44** | -0.52*** |
|  | (0.39) | (1.62) | (-1.58) | (-0.01) | (1.12) | (-1.93) | (-0.48) | (2.11) | (-2.99) |
| MOM | -0.24*** | -0.72*** | 0.48*** | -0.25*** | -0.46*** | 0.21*** | 0.08 | -0.75*** | 0.83*** |
|  | (-3.04) | (-7.11) | (5.57) | (-3.17) | (-6.12) | (4.25) | (1.38) | (-8.04) | (10.77) |
| RMW | -0.11 | 0.36 | -0.47** | -0.10 | 0.21 | -0.31** | -0.34** | 0.51** | -0.85*** |
|  | (-0.51) | (1.39) | (-2.14) | (-0.49) | (0.97) | (-2.33) | (-2.07) | (2.05) | (-4.49) |
| CMA | 0.07 | -0.23 | 0.30 | -0.10 | -0.15 | 0.04 | -0.04 | -0.28 | 0.24 |
|  | (0.30) | (-0.90) | (1.49) | (-0.52) | (-0.65) | (0.32) | (-0.20) | (-1.01) | (1.01) |
| Observations | 2,579 | 2,579 | 2,579 | 2,579 | 2,579 | 2,579 | 2,579 | 2,579 | 2,579 |
| Adjusted $R^2$ | 0.33 | 0.42 | 0.08 | 0.38 | 0.42 | 0.04 | 0.35 | 0.43 | 0.23 |

|  | NN | | | Comb | | | OLS | | |
|---|---|---|---|---|---|---|---|---|---|
|  | (10) | (11) | (12) | (13) | (14) | (15) | (16) | (17) | (18) |
|  | Top | Bottom | T-B | Top | Bottom | T-B | Top | Bottom | T-B |
| Alpha | 8.12*** | 3.57 | 4.55** | 8.57*** | 4.45* | 4.11*** | 4.64* | 6.63*** | -1.99 |
|  | (3.76) | (1.40) | (2.14) | (4.02) | (1.85) | (2.70) | (1.85) | (2.90) | (-0.90) |
| MKT | 0.73*** | 0.80*** | -0.08* | 0.76*** | 0.87*** | -0.12*** | 0.83*** | 0.79*** | 0.04 |
|  | (12.39) | (13.15) | (-1.78) | (13.43) | (14.14) | (-4.19) | (10.85) | (14.78) | (0.88) |
| SMB | -0.63*** | -0.58*** | -0.06 | -0.65*** | -0.57*** | -0.08 | -0.62*** | -0.56*** | -0.06 |
|  | (-7.51) | (-5.14) | (-0.72) | (-6.96) | (-5.22) | (-1.44) | (-4.69) | (-6.61) | (-0.66) |
| HML | -0.16 | 0.19 | -0.35*** | -0.04 | 0.29 | -0.33*** | 0.11 | 0.04 | 0.07 |
|  | (-0.92) | (1.03) | (-2.73) | (-0.26) | (1.58) | (-3.24) | (0.48) | (0.21) | (0.40) |
| MOM | -0.07 | -0.70*** | 0.64*** | -0.12* | -0.66*** | 0.54*** | -0.18* | -0.47*** | 0.29* |
|  | (-0.85) | (-6.80) | (8.68) | (-1.81) | (-7.36) | (9.96) | (-1.94) | (-3.92) | (1.88) |
| RMW | 0.04 | 0.18 | -0.14 | -0.13 | 0.32 | -0.44*** | -0.07 | 0.15 | -0.22 |
|  | (0.20) | (0.75) | (-0.81) | (-0.69) | (1.37) | (-3.60) | (-0.28) | (0.57) | (-0.68) |
| CMA | 0.04 | -0.27 | 0.31* | -0.01 | -0.23 | 0.22 | -0.25 | 0.09 | -0.34* |
|  | (0.17) | (-1.14) | (1.80) | (-0.05) | (-0.98) | (1.62) | (-0.91) | (0.45) | (-1.80) |
| Observations | 2,579 | 2,579 | 2,579 | 2,579 | 2,579 | 2,579 | 2,579 | 2,579 | 2,579 |
| Adjusted $R^2$ | 0.35 | 0.40 | 0.13 | 0.38 | 0.45 | 0.19 | 0.36 | 0.39 | 0.02 |

This table presents the daily performance of portfolios formed on return forecasts by foreign signals. To predict the daily returns of the S&P 500 stocks, we use foreign signals and various machine learning algorithms, least absolute shrinkage and selection operator (LASSO) regression, random forest (RF), gradient boosted tree (GB), neural network (NN), and ordinary least squares regression (OLS). Foreign signals are lagged weekly returns of leading stock market indices in 47 non-U.S. markets during the previous four weeks. Top refers to the portfolios of stocks with top 5% return forecasts. Bottom means the portfolios of stocks with bottom 5% return forecasts. We equally weight stocks in each portfolio. T-B is the portfolio that buys stocks with top 5% return forecasts and sells short stocks with bottom 5% return forecasts. Portfolios are rebalanced every day. We report the intercept (Alpha) and loadings from regressions of Top or Bottom portfolio returns in excess of the daily one-month U.S. Treasury yield on the Fama and French (2017) global ex-U.S. factors. We also report the intercept (Alpha) and loadings from the regression of T-B portfolio returns on the same set of factors. In columns (13) to (15), we equally weight the returns of portfolios based on LASSO, RF, GB, and NN for Top, Bottom, or T-B, respectively. All results are in basis points. Robust *t*-statistics are reported in parentheses. *, **, and *** denote statistical significance at the 10%, 5%, and 1% level, respectively.

**Table A10: Monthly Performance of Portfolios Formed on Return Forecasts by Foreign Signals**

| | LASSO | | | RF | | | GB | | |
|---|---|---|---|---|---|---|---|---|---|
| | (1) | (2) | (3) | (4) | (5) | (6) | (7) | (8) | (9) |
| | Top | Bottom | T-B | Top | Bottom | T-B | Top | Bottom | T-B |
| Mean return | 0.0213*** | 0.0097 | 0.0116*** | 0.0198*** | 0.0119** | 0.0079** | 0.0199*** | 0.0121* | 0.0078 |
| | (3.80) | (1.47) | (2.73) | (4.15) | (2.20) | (2.45) | (4.54) | (1.81) | (1.62) |
| | | | | | | | | | |
| Alpha | 0.0067** | -0.0041 | 0.0108** | 0.0078*** | -0.0009 | 0.0087*** | 0.0082*** | -0.0011 | 0.0093*** |
| | (2.54) | (-1.33) | (2.60) | (3.98) | (-0.41) | (2.75) | (3.63) | (-0.41) | (2.62) |
| MKT | 1.2129*** | 1.2148*** | -0.0018 | 1.0353*** | 1.0985*** | -0.0631 | 1.0054*** | 1.1621*** | -0.1567 |
| | (10.89) | (11.33) | (-0.02) | (19.70) | (20.40) | (-0.80) | (13.42) | (8.99) | (-0.97) |
| SMB | 0.3813*** | 0.2760* | 0.1053 | 0.3006*** | 0.2233** | 0.0773 | 0.2425** | 0.4213*** | -0.1788 |
| | (2.88) | (1.76) | (0.49) | (2.86) | (2.34) | (0.46) | (2.33) | (2.96) | (-0.94) |
| HML | 0.0231 | 0.2231 | -0.2000 | 0.0697 | 0.1641 | -0.0944 | 0.2445*** | 0.2777* | -0.0332 |
| | (0.17) | (1.19) | (-0.80) | (0.74) | (1.46) | (-0.62) | (2.68) | (1.73) | (-0.17) |
| MOM | -0.1646 | -0.6529*** | 0.4883*** | -0.2013** | -0.4093*** | 0.2080 | 0.1024 | -0.6320*** | 0.7344*** |
| | (-1.52) | (-6.01) | (2.78) | (-2.59) | (-4.71) | (1.63) | (1.24) | (-5.74) | (4.43) |
| RMW | 0.0106 | 0.0063 | 0.0043 | -0.1848 | 0.0126 | -0.1974 | -0.2348 | 0.1652 | -0.3999 |
| | (0.06) | (0.03) | (0.02) | (-1.46) | (0.08) | (-0.88) | (-1.59) | (0.78) | (-1.55) |
| CMA | 0.1494 | 0.2514 | -0.1020 | 0.0553 | 0.1271 | -0.0718 | -0.1343 | 0.1841 | -0.3184 |
| | (0.73) | (0.99) | (-0.29) | (0.34) | (0.78) | (-0.32) | (-0.77) | (0.85) | (-1.13) |
| | | | | | | | | | |
| Observations | 123 | 123 | 123 | 123 | 123 | 123 | 123 | 123 | 123 |
| Adjusted $R^2$ | 0.7653 | 0.8140 | 0.1627 | 0.8338 | 0.8374 | 0.0564 | 0.7738 | 0.8090 | 0.3885 |

|  | NN | | | Comb | | | OLS | | |
|---|---|---|---|---|---|---|---|---|---|
|  | (10) | (11) | (12) | (13) | (14) | (15) | (16) | (17) | (18) |
|  | Top | Bottom | T-B | Top | Bottom | T-B | Top | Bottom | T-B |
| Mean return | 0.0199*** | 0.0072 | 0.0126*** | 0.0203*** | 0.0103* | 0.0100*** | 0.0126** | 0.0151*** | -0.0025 |
|  | (4.49) | (1.22) | (3.06) | (4.50) | (1.72) | (3.07) | (2.20) | (2.71) | (-0.50) |
| Alpha | 0.0055** | -0.0054** | 0.0109*** | 0.0071*** | -0.0029 | 0.0099*** | -0.0025 | 0.0033 | -0.0058 |
|  | (2.54) | (-2.09) | (3.08) | (4.45) | (-1.33) | (3.79) | (-0.78) | (1.09) | (-1.20) |
| MKT | 1.1093*** | 1.1431*** | -0.0338 | 1.0903*** | 1.1543*** | -0.0641 | 1.2387*** | 1.0367*** | 0.2020 |
|  | (13.18) | (13.14) | (-0.29) | (21.24) | (15.30) | (-0.79) | (10.87) | (10.04) | (1.14) |
| SMB | 0.2182** | 0.1747* | 0.0435 | 0.2858*** | 0.2748*** | 0.0110 | 0.2931* | 0.1712 | 0.1218 |
|  | (2.10) | (1.66) | (0.29) | (3.57) | (2.75) | (0.08) | (1.80) | (1.08) | (0.44) |
| HML | -0.1297 | 0.1104 | -0.2401 | 0.0495 | 0.1941 | -0.1446 | 0.2431 | 0.0541 | 0.1890 |
|  | (-1.04) | (0.88) | (-1.26) | (0.61) | (1.58) | (-0.96) | (0.90) | (0.30) | (0.48) |
| MOM | 0.0963 | -0.5706*** | 0.6670*** | -0.0430 | -0.5674*** | 0.5245*** | -0.0335 | -0.4629*** | 0.4294 |
|  | (0.73) | (-5.17) | (4.30) | (-0.52) | (-6.91) | (4.13) | (-0.20) | (-3.40) | (1.53) |
| RMW | 0.1383 | -0.2675 | 0.4059* | -0.0690 | -0.0208 | -0.0482 | 0.1058 | -0.2668 | 0.3726 |
|  | (0.92) | (-1.56) | (1.69) | (-0.69) | (-0.14) | (-0.27) | (0.57) | (-1.37) | (1.17) |
| CMA | 0.4427** | 0.0624 | 0.3803 | 0.1323 | 0.1559 | -0.0236 | -0.0003 | 0.3525 | -0.3528 |
|  | (2.28) | (0.32) | (1.34) | (0.96) | (0.93) | (-0.11) | (-0.00) | (1.38) | (-0.75) |
| Observations | 123 | 123 | 123 | 123 | 123 | 123 | 123 | 123 | 123 |
| Adjusted $R^2$ | 0.7671 | 0.8152 | 0.3399 | 0.8783 | 0.8760 | 0.3644 | 0.7166 | 0.6977 | 0.0222 |

This table presents the month performance of portfolios formed on foreign signal return forecasts. We aggregate the daily portfolio returns in Table 1 to monthly returns. Top refers to the portfolios of stocks with top 5% return forecasts. Bottom indicates the portfolios of stocks with bottom 5% return forecasts. We equally weight stocks in each portfolio. T-B is the portfolio that buys stocks with top 5% return forecasts and sells short stocks with bottom 5% return forecasts. We report the mean return and intercept (Alpha) and loadings from regressions of Top or Bottom portfolio returns in excess of the daily one-month U.S. Treasury yield on the Fama and French (1993, 2015) MKT, SMB, HML, RMW, CMA, and Carhart (1997) MOM factors. We also report the mean return and intercept (Alpha) and loadings from the regression of T-B portfolio returns on the same set of factors. In columns (13) to (15), we equally weight the returns of portfolios based on LASSO, RF, GB, and NN for Top, Bottom, or T-B, respectively. All results are based on monthly returns. Robust *t*-statistics are reported in parentheses. *, **, and *** denote statistical significance at the 10%, 5%, and 1% level, respectively.

Table A11: Daily Performance of Portfolios Formed on Return Forecasts by Foreign Signals, Value-weighted

|  | LASSO | | | RF | | | GB | | |
|---|---|---|---|---|---|---|---|---|---|
|  | (1) | (2) | (3) | (4) | (5) | (6) | (7) | (8) | (9) |
|  | Top | Bottom | T-B | Top | Bottom | T-B | Top | Bottom | T-B |
| Mean return | 9.21*** | 5.14 | 4.07 | 7.46*** | 8.16*** | -0.70 | 9.85*** | 6.78** | 3.07 |
|  | (3.05) | (1.57) | (1.54) | (2.85) | (3.08) | (-0.38) | (3.75) | (2.27) | (1.22) |
|  |  |  |  |  |  |  |  |  |  |
| Alpha | 2.24 | -1.65 | 3.89 | 0.99 | 1.79 | -0.80 | 3.15** | 0.63 | 2.52 |
|  | (1.40) | (-1.05) | (1.64) | (0.80) | (1.51) | (-0.45) | (2.55) | (0.50) | (1.37) |
| MKT | 1.18*** | 1.21*** | -0.03 | 1.09*** | 1.08*** | 0.01 | 1.09*** | 1.09*** | -0.00 |
|  | (46.81) | (38.43) | (-0.55) | (49.74) | (58.85) | (0.16) | (49.55) | (49.85) | (-0.03) |
| SMB | 0.18*** | 0.07* | 0.11* | -0.00 | 0.05* | -0.05 | -0.02 | 0.12*** | -0.14*** |
|  | (3.85) | (1.94) | (1.90) | (-0.11) | (1.78) | (-1.29) | (-0.63) | (3.47) | (-2.97) |
| HML | 0.22*** | 0.08** | 0.14** | 0.06* | 0.07** | -0.01 | 0.09** | 0.12*** | -0.04 |
|  | (4.05) | (2.04) | (2.04) | (1.84) | (2.57) | (-0.19) | (2.28) | (3.36) | (-0.77) |
| MOM | 0.09** | -0.58*** | 0.67*** | -0.02 | -0.26*** | 0.24*** | 0.26*** | -0.58*** | 0.84*** |
|  | (2.48) | (-12.94) | (9.75) | (-0.47) | (-12.61) | (5.49) | (11.35) | (-20.58) | (23.91) |
| RMW | -0.04 | -0.24*** | 0.20** | -0.04 | -0.04 | 0.00 | -0.04 | -0.11*** | 0.07 |
|  | (-0.77) | (-4.24) | (2.18) | (-1.03) | (-1.12) | (0.05) | (-1.09) | (-2.79) | (1.17) |
| CMA | -0.16* | 0.10 | -0.26** | -0.18*** | 0.07 | -0.25*** | -0.11* | 0.07 | -0.17** |
|  | (-1.94) | (1.36) | (-2.25) | (-2.95) | (1.36) | (-3.18) | (-1.66) | (1.11) | (-1.99) |
|  |  |  |  |  |  |  |  |  |  |
| Observations | 2,579 | 2,579 | 2,579 | 2,579 | 2,579 | 2,579 | 2,579 | 2,579 | 2,579 |
| Adjusted $R^2$ | 0.72 | 0.77 | 0.19 | 0.78 | 0.80 | 0.09 | 0.78 | 0.82 | 0.46 |

|  | NN | | | Comb | | | OLS | | |
|---|---|---|---|---|---|---|---|---|---|
|  | (10) | (11) | (12) | (13) | (14) | (15) | (16) | (17) | (18) |
|  | Top | Bottom | T-B | Top | Bottom | T-B | Top | Bottom | T-B |
| Mean return | 9.33*** | 2.61 | 6.71*** | 8.96*** | 5.68** | 3.29* | 6.97** | 6.55** | 0.42 |
|  | (3.54) | (0.90) | (2.80) | (3.55) | (2.05) | (1.88) | (2.57) | (2.51) | (0.18) |
| Alpha | 2.50* | -3.47** | 5.97*** | 2.22** | -0.67 | 2.89** | 0.46 | 0.72 | -0.26 |
|  | (1.87) | (-2.42) | (2.92) | (2.43) | (-0.73) | (2.20) | (0.32) | (0.50) | (-0.11) |
| MKT | 1.10*** | 1.07*** | 0.03 | 1.11*** | 1.11*** | 0.00 | 1.09*** | 1.00*** | 0.09*** |
|  | (53.12) | (52.02) | (0.87) | (68.63) | (73.22) | (0.04) | (56.39) | (46.44) | (2.74) |
| SMB | -0.05 | 0.07** | -0.12** | 0.03 | 0.08*** | -0.05 | 0.04 | 0.11*** | -0.06 |
|  | (-1.47) | (2.16) | (-2.30) | (0.96) | (3.46) | (-1.56) | (1.37) | (3.28) | (-1.24) |
| HML | -0.10*** | 0.09** | -0.18*** | 0.07** | 0.09*** | -0.02 | 0.09** | 0.07* | 0.03 |
|  | (-2.69) | (2.21) | (-3.27) | (2.21) | (3.94) | (-0.65) | (2.41) | (1.76) | (0.44) |
| MOM | 0.08* | -0.47*** | 0.55*** | 0.10*** | -0.47*** | 0.58*** | -0.02 | -0.19*** | 0.17 |
|  | (1.69) | (-8.95) | (13.28) | (4.48) | (-16.20) | (20.87) | (-0.44) | (-2.71) | (1.51) |
| RMW | 0.03 | -0.18*** | 0.21*** | -0.02 | -0.14*** | 0.12*** | -0.08* | -0.11** | 0.03 |
|  | (0.68) | (-3.88) | (2.88) | (-0.80) | (-4.91) | (2.79) | (-1.67) | (-2.02) | (0.36) |
| CMA | 0.03 | 0.06 | -0.03 | -0.11** | 0.07* | -0.18*** | 0.01 | 0.04 | -0.03 |
|  | (0.51) | (0.90) | (-0.33) | (-2.15) | (1.78) | (-3.02) | (0.12) | (0.49) | (-0.26) |
| Observations | 2,579 | 2,579 | 2,579 | 2,579 | 2,579 | 2,579 | 2,579 | 2,579 | 2,579 |
| Adjusted $R^2$ | 0.75 | 0.76 | 0.28 | 0.87 | 0.89 | 0.44 | 0.72 | 0.71 | 0.02 |

This table presents the daily performance of portfolios formed on return forecasts by foreign signals. To predict the daily returns of the S&P 500 stocks, we use foreign signals and various machine learning algorithms, least absolute shrinkage and selection operator (LASSO) regression, random forest (RF), gradient boosted tree (GB), neural network (NN), and ordinary least squares regression (OLS). Foreign signals are lagged weekly returns of leading stock market indices in 47 non-U.S. markets during the previous four weeks. Top refers to the portfolios of stocks with top 5% return forecasts. Bottom means the portfolios of stocks with bottom 5% return forecasts. We value weight stocks in each portfolio by their market cap at the end of the previous trading day. T-B is the portfolio that buys stocks with top 5% return forecasts and sells short stocks with bottom 5% return forecasts. Portfolios are rebalanced every day. We report the mean return and intercept (Alpha) and loadings from regressions of Top or Bottom portfolio returns in excess of daily one-month U.S. Treasury yield on the Fama and French (1993, 2015) MKT, SMB, HML, RMW, CMA, and Carhart (1997) MOM factors. We also report the mean return and intercept (Alpha) and loadings from the regression of T-B portfolio returns on the same set of factors. In columns (13) to (15), we equally weight the returns of portfolios based on LASSO, RF, GB, and NN for Top, Bottom, or T-B, respectively. All results are in basis points. Robust *t*-statistics are reported in parentheses. *, **, and *** denote statistical significance at the 10%, 5%, and 1% level, respectively.

## Table A12: Elastic Net Analysis

Panel A: Daily Portfolio Returns

|  | LASSO | | |
| --- | --- | --- | --- |
|  | (1) Top | (2) Bottom | (3) T-B |
| Mean return | 10.11*** | 4.74 | 5.36** |
|  | (3.36) | (1.39) | (2.35) |
| Alpha | 3.41** | -2.02 | 5.42*** |
|  | (2.44) | (-1.42) | (2.66) |
| Factors | Y | Y | Y |
| Observations | 2,579 | 2,579 | 2,579 |
| Adjusted $R^2$ | 0.78 | 0.82 | 0.18 |

Panel B: Out-of-sample $R^2_{oos}$

|  | Fraction of Stocks with $R^2_{oos}>0$ | $R^2_{oos}$ Distribution | | | | | | | | Mean (Stocks with $R^2_{oos}>0$) |
| --- | --- | --- | --- | --- | --- | --- | --- | --- | --- | --- |
|  |  | 99th | 95th | 90th | 75th | 50th | 25th | 10th | 5th |  |
| Elastic Net | 24.4% | 5.1% | 2.6% | 1.5% | 0.0% | -0.3% | -2.1% | -5.4% | -9.1% | 1.6% |

This table presents the results based on Elastic Net. Foreign signals are lagged weekly returns of leading stock market indices in 47 non-U.S. markets during the previous four weeks. Panel A presents the daily performance of portfolios formed on foreign signal return forecasts. Top refers to the portfolios of stocks with top 5% return forecasts. Bottom indicates the portfolios of stocks with bottom 5% return forecasts. We equally weight stocks in each portfolio. T-B is the portfolio that buys stocks with top 5% return forecasts and sells short stocks with bottom 5% return forecasts. Portfolios are rebalanced every day. We report the mean return and intercept (Alpha) from regressions of Top or Bottom portfolio returns in excess of the daily one-month U.S. Treasury yield on the Fama and French (1993, 2015) MKT, SMB, HML, RMW, CMA, and Carhart (1997) MOM factors. We also report the mean return and intercept (Alpha) and loadings from the regression of T-B portfolio returns on the same set of factors. For brevity, we omit the loadings on factors. Robust t-statistics are reported in parentheses. Panel B reports the out-of-sample $R^2_{oos}$.

# Internet Appendix II

## The Leading Stock Market Indices of 47 non-U.S. Markets

| Market | Leading stock market index | Market | Leading stock market index |
|---|---|---|---|
| Australia | S&P/ASX 200 | Malaysia | FTSE BURSA Malaysia KLCI |
| Austria | Austrian Traded Index | Mexico | Mexico IPC (BOLSA) |
| Belgium | BEL 20 | Netherlands | AEX Index (AEX) |
| Brazil | Bovespa Index | New Zealand | S&P/NZX 50 |
| Canada | S&P/TSX Composite Index | Norway | OSLO SE OBX |
| Chile | S&P IPSA CLP Index | Peru | S&P/BVL General (IGBVL) |
| China | Shanghai Composite Index | Philippines | Philippine Stock Exchange Composite Index (PSEi) |
| Colombia | COLCAP | Poland | Warsaw General Index |
| Czech Republic | Prague PX index | Portugal | Portugal PSI-20 |
| Denmark | OMXC20 | Qatar | QE All Share Index |
| Egypt | EGX30 | Russia | MOEX Russia Index |
| Finland | OMX Helsinki 25 (OMXH25) | Saudi Arabia | Saudi Tadawul All Share (TASI) |
| France | France CAC 40 | Singapore | Straits Times Index |
| Germany | DAX | South Africa | FTSE/JSE All Share |
| Greece | Athex Composite | South Korea | Korea Composite (KOSPI) |
| Hong Kong (China) | Hang Seng Index | Spain | IBEX 35 |
| Hungary | Budapest Stock Exchange Index (BUX) | Sweden | OMX Stockholm 30 (OMXS30) |
| India | S&P BSE SENSEX | Switzerland | Swiss Market Index (SMI) |
| Indonesia | IDX Composite | Taiwan (China) | Taiwan Capitalization Weighted Stock Index |
| Ireland | ISEQ All Share Index | Thailand | Bangkok S.E.T. |
| Israel | TA 35 | Turkey | BIST National 100 |
| Italy | FTSE MIB Index | United Arab Emirates | Dubai Financial Market Index |
| Japan | NIKKEI 225 | United Kingdom | FTSE 100 |
| Kuwait | S&P Kuwait BMI | | |

# Internet Appendix III

# Hyperparameter Tuning

Hyperparameters are the parameters used to control the learning process of machine learning algorithms. We tune hyperparameters using *Optuna* (Akiba et al. 2019). *Optuna* is an automatic hyperparameter optimization software framework, particularly designed for machine learning. We tune hyperparameters based on randomly picked 2,000 stock-quarters from 2002Q1 to 2011Q4. Following the model training and predicting procedures described in the paper, we use *Optuna* to identify the hyperparameter values that yield the lowest predictive errors to forecast daily returns for each of these 2,000 stock-quarters. The median hyperparameter values from these 2,000 independent searches are used for our training and forecasting for the sample from 2012Q1 to 2022Q1.

We train least absolute shrinkage and selection operator (LASSO), random forest (RF), and gradient boosted tree (GB) using *Scikit-learn* library and train neural network (NN) using *Tensorflow*. We detail the hyperparameters that we tune as follows. We follow the names of the hyperparameters used by *Scikit-learn* and *Tensorflow*.

| Algorithm | Hyperparameter | Hyperparameter description |
|---|---|---|
| LASSO | *alpha* | a hyperparameter controlling the regularization strength |
| RF | *n_estimators* | the number of trees in the forest |
| | *max_depth* | the maximum depth of each tree |
| | *min_samples_leaf* | the minimum number of samples required to be at a leaf node |
| | *max_samples* | the number of observations drawn from the training sample to train each tree |
| | *max_features* | the number of features to consider when looking for the best split |
| GB | *n_estimators* | the number of boosting stages to perform |
| | *max_depth* | the maximum depth of each tree |

|     | *min_samples_leaf* | the minimum number of samples required to be at a leaf node |
|     | *learning_rate* | a hyperparameter controlling the contribution of each tree |
|     | *subsample* | the fraction of the training sample used to train each tree |
|     | *max_features* | the number of features to consider when looking for the best split |
| NN  | *epochs* | the number of epochs to train the model. An epoch is an iteration over the entire training sample. |
|     | *batch_size* | the number of samples per batch of computation |
|     | *No. of layers* | the number of layers in the neural network besides the first and final layers |
|     | *No. of neurons* | the number of neurons in each layer besides the final layer |
|     | *learning_rate* | the learning rate used by the Adam optimizer |
|     | *L1* | a hyperparameter controlling the strength of L1 kernel regularization |